\documentclass[
aps,%
12pt,%
final,%
notitlepage,%
oneside,%
onecolumn,%
nobibnotes,%
nofootinbib,%
superscriptaddress,%
noshowpacs,%
centertags,%
natbib
]{revtex4}

\usepackage{epstopdf}
\usepackage[usenames]{color}
\usepackage{graphicx,times}
\usepackage{natbib}
\usepackage{amssymb,amsmath}

\usepackage{xcolor}
\usepackage{lscape}
\usepackage{longtable}
\usepackage{array}
\usepackage{lscape}
\usepackage{xtab}

\begin{document}
\newcommand{\xh}{X_{\rm H}}
\newcommand{\eg}{E_{\rm gap}}
\newcommand{\nc}{N_{\rm C}}
\newcommand{\ach}{\hbox{a-C:H}}
\newcommand{\acet}{\hbox{C$_2$}}
\newcommand{\qpah}{\hbox{$q_{\rm PAH}$}}
\newcommand{\sigmach}{\sigma^{\rm CH}_{\rm loss}}
\newcommand{\na}{N^{\rm a}}
\newcommand{\nbineg}{N^{\rm eg}}

\def\apjl{ApJL}
\def\apjs{ApJS}
\def\aap{A\&A}
\def\mnras{MNRAS}
\def\pasp{PASP}
\def\nar{New Astron. Reviews}
\def\aaps{Astron. Astroph. Suppl.}
\def\aj{AJ}
\def\araa{Ann.Rev.Astron.Astrophys}
\def\pasj{PASJ}
\title{Studies of Star-forming Complexes in the Galaxies NGC 628, NGC 2976, and NGC 3351}

\author{\firstname{K.~I.}~\surname{Smirnova}}
\email[E-mail: ]{Arashu@rambler.ru}
\affiliation{Ural Federal University, Ekaterinburg, Russia}

\author{\firstname{D.~S.}~\surname{Wiebe}}
\affiliation{Institute of Astronomy, Russian Academy of Sciences, Moscow, Russia}

\begin{abstract}
We analyze parameters of the interstellar matter emission in star-forming complexes in the high metallicity galaxies NGC~628, NGC~2976, and NGC~3351, which have different morphological types. The relation between H$\alpha$ emission and emission in CO and HI lines is considered along with the relation between H$\alpha$ emission and dust emission in the infrared range (IR). The fluxes and surface brightnesses in the UV and IR correlate well with H$\alpha$ emission. The HI emission also correlates well with H$\alpha$, while the correlation between the CO and H$\alpha$ emission is much less prominent. The ratio of the fluxes at 8 and 24 $\mu$m decreases with increasing H$\alpha$ flux. This may be due to changes in the properties of the dust ensemble (a decrease in the mass fraction of polycyclic aromatic hydrocarbons) or to changes in excitation conditions. Analysis of the kinematics of the CO lines shows that the CO flux grows with increasing velocity scatter $\Delta V$ when $\Delta V\lesssim70$~km/s. Preliminary evidence for the existence of star-forming complexes with higher values of $\Delta V$ is presented, and the increase in the velocity scatter is accompanied by a decrease in the CO line luminosity of the complex.

\end{abstract}

\maketitle

\section{INTRODUCTION}
Star formation is a key process in the Universe, which brings about a gradual transformation of gas into compact objects with (potentially) unlimited lifetimes (degenerate dwarfs, neutron stars, black holes, planets). Studies of star formation have always attracted considerable interest, but possibilities for such studies have expanded appreciably recently, thanks to observations of the birth of stars in galaxies with parameters significantly different from those of the Milky Way. Studies of extragalactic star-forming regions and complexes enable the identification of various metrics of the star formation rate (SFR), and can make it possible to find links between the SFR and various parameters of the interstellar medium (ISM) in these galaxies, both on a global scale (for the galaxy as a whole) and on smaller spatial scales.

In practice, the most direct way to determine the SFR is to count the number of stars and/or young stellar objects in various age ranges \cite{1998ARA&A..36..189K,2000A&A...358..869R,2010ApJ...710L..11R}.However,this method is demanding in terms of both the quality and quantity of the observational data required, making its application limited to our Galaxy or other very nearby galaxies. Otherwise, we must rely on indirect metrics of the SFR. A conceptual average SFR over a time of the order of a Gyr can be obtained from multicolor photometry. However, more detailed data are necessary for estimating the current (``instantaneous'') SFR.

Metrics of the SFR are usually directly or indirectly associated with the UV radiation of massive stars with the shortest lifetimes \cite{2013seg..book..419C}. Such metrics include, for instance, observations of the UV continuum or the H$\alpha$ line (assuming that the UV radiation of massive stars is reprocessed into H$\alpha$ emission in zones and complexes of ionized hydrogen). However, in this case we face the problem of dust, which absorbs a fraction of the UV radiation and H$\alpha$ emission. This can make conclusions on the SFR less reliable, especially for galaxies at high redshifts. Observations of the infrared (IR) and submillimeter emission of dust heated by the shorter-wavelength radiation of young stars could be an alternative, but in this case the opposite problem arises: the dust does not absorb and re-emit all the radiation of young stars, and the relationship between the SFR and the IR luminosity can be specific to a given galaxy, depending on its metallicity, the characteristics of its dust distribution, etc. In addition, evolved stellar populations can make a significant contribution to heating of dust. If emission bands of polycyclic aromatic hydrocarbons (PAHs) are used as a metric of the SFR, additional difficulties arise, due to the fact that the abundance of PAHs in star-forming regions may change with time \cite{2014AstL...40..278W,2014MNRAS.444..757K}.

To ensure good accuracy, it is advisable to use several metrics that compensate for each others' shortcomings. One example of a comparative analysis using various metrics of star formation is presented by Bendo et al. \cite{2017Bendo}, who compared the SFR in the galaxy NGC 5253 estimated from the H$\alpha$30 radiorecombination line in a star-forming zone (such estimates should be the least susceptible to various interferences) with the SFRs derived from other traditional metrics. It turned out that these various estimates of the SFR differed significantly. Bendo et al. \cite{2017Bendo} suggested that this was due to the low metallicity of the galaxy, and the fact that, for some reason, the dust in this galaxy was heated to a higher temperature than in comparable systems with the ``usual'' metallicity.

Another recent example of a comparison of SFR estimates obtained using different metrics is \cite{2018Audcent-Ross}, where two metrics were used to determine the density of the SFR---the emission in the H$\alpha$ line and the far-UV, with their sample containing an appreciable number of dwarf galaxies. It turned out that the ratio of these two fluxes in different galaxies displayed both significant scatter and systematic variations. This means that the estimates of the SFR for a specific galaxy derived from H$\alpha$ and UV observations can differ significantly, although this difference almost disappears when the global density of the SFR in the local Universe is considered.

Overall, it is obvious that the use of a single SFR indicator can lead to substantial errors in estimates \cite{2010Calzetti}. On the other hand, differences in the SFRs estimated using different indicators may point toward important differences in the parameters of star formation processes under different conditions, such as variations in the initial mass function for massive stars (which are usually responsible for the action of one of the indicators). Therefore, comparative studies with a larger number of SFR indicators, encompassing the entire range of physical conditions in the ISM and systems having a wide range of metallicities, are necessary.

In our previous study \cite{ourpreviouswork}, we considered star-forming complexes (SFCs) in 11 galaxies that were included in a number of surveys, including THINGS \cite{THINGS} (the 21-cm HI line), KINGFISH \cite{kingfish} (far-IR at 70, 100, and 160 $\mu$m with the Herschel Space Telescope), SINGS \cite{sings} (near- and mid-IR bands at 3.6, 4.5, 5.8, 8.0, and 24 $\mu$m with the Spitzer Space Telescope), and HERACLES \cite{heracles} (CO (2–1) line using the 30-m IRAM telescope). We investigated the relationship between different components of the ISM in particular SFCs. Infrared data were used to estimate the total mass of dust ($M$), the mass fraction of PAHs ($q_{\rm PAH}$), and the average radiation intensity $U_{\min}$ using the model of \cite{2007Draine}. In our present study, we explore three galaxies from the sample \cite{ourpreviouswork} for which homogeneous archival observational H$\alpha$ data are available. We discuss these data in detail and compare them with the data obtained in \cite{ourpreviouswork}.

A significant role in assessing and understanding the distributions of certain components of the ISM and the stability of the gaseous disks of galaxies is played by the gas velocity dispersion. As a rule, this is measured using observations in the 21 cm HI line and CO lines. Due to the greater availability of data in the 21 cm line, studies of its dispersion are more common. To date, there have been several studies where the velocity dispersions in these two lines have been compared. For example, the velocity dispersions for the atomic and molecular gas in 12 spiral galaxies were compared in \cite{2013Caldu-Primo}, leading to the conclusion that these dispersions are similar on scales of the order of several kpc. This contradicts a picture in which the thickness of the molecular disk is significantly (by a factor of a few) smaller than the thickness of the atomic disk, and can be explained by the presence of a thick disk of molecular gas comparable to the disk of neutral hydrogen. A thick molecular disk was discovered in the galaxy M51 in \cite{2013Pety} through observations in the CO (1---0) line, which made it possible for the first time to reconstruct the large scale distribution of molecular gas with a resolution of 40 pc.

A similar conclusion about the presence of molecular gas with a high velocity dispersion in galaxies was also drawn in \cite{2016Mogotsi}, where it is shown that the dispersion of the molecular gas is about a factor of 1.5 lower than the velocity dispersion of the atomic gas in regions of high CO brightness, but the width of the CO lines increases in the transition to molecular gas with lower brightness. Mogotsi et al. \cite{2016Mogotsi} concluded that disk galaxies contain not only a thin disk with a high CO density and relatively low velocities, but also a weaker, high-velocity diffuse component of the molecular disk.

In our present study, we complement analyses of the IR and H$\alpha$ emission of extragalactic SFRs by considering the velocity dispersions in CO lines. Unlike \cite{2016Mogotsi}, we consider not only complexes with bright CO emission, but also those that are prominent sources of IR and/or H$\alpha$ emission, but may not be bright CO sources.

The selected galaxies have different morphological types. NGC~3351 is a ring galaxy with a bar; the youth of the SFCs in this galaxy has been noted in various studies (see, e.g., \cite{Swartz2006, Elmegreen1997}). NGC~2976 is a dwarf galaxy populated by both equally distributed old populations and young SFCs in a region of the galactic disk with a radius of about 3 kpc \cite{2013Gusev}. This galaxy displays a tendency for the ages of stars to increase with galactocentric distance. NGC~628 is a classic spiral galaxy viewed almost side-on, enabling us to isolate the largest number of SFCs. A homogeneous distribution of stellar complexes is observed along its long arm, while this distribution is not homogeneous along its short arm \cite{2013Gusev}. This does not have a unique explanation, but it would seem to be unusual for a seemingly ordinary spiral galaxy. The stellar population of NGC~628 is mostly old. However, there are small ``islets'' of SFCs formed relatively recently. Figure 7 of \cite{2014Sanchez-Blazques} shows two-dimensional maps of the age distribution of the stars, based on their masses and luminosities.

\section{THE OBSERVATIONS AND THEIR REDUCTION}
We used the following archival observations in our study. The data in the near- and mid-IR (3.6, 4.5, 5.8, 8.0, and 24 $\mu$m) were taken from the Spitzer SINGS survey\footnote{http:/sings.stsci.edu} \cite{sings}. The far-IR data (70, 100, and 160 $\mu$m) were taken from the Herschel KINGFISH survey\footnote{http://herschel.esac.esa.int/Science\_Archive.shtml} \cite{kingfish}]. The 21 cm line observations were taken from the THINGS survey\footnote{http://www.mpia.de/THINGS/Data.html} \cite{THINGS}. The CO (2–1) line observations (a molecular-hydrogen indicator) were taken from the HERACLES survey\footnote{http://www.cv.nrao.edu/leroy/heracles\_data/}\cite{heracles}. The H$\alpha$ images of the galaxies were obtained using the 2.3-meter BOK telescope of the Steward Observatory \cite{2008ApJS..178..247K}. Finally, observations by the GALEX telescope \cite{2005ApJ...619L...1M} were obtained from the MAST archive\footnote{http://archive.stsci.edu}.

\begin{figure*}[t!]
\includegraphics[width=0.3\textwidth]{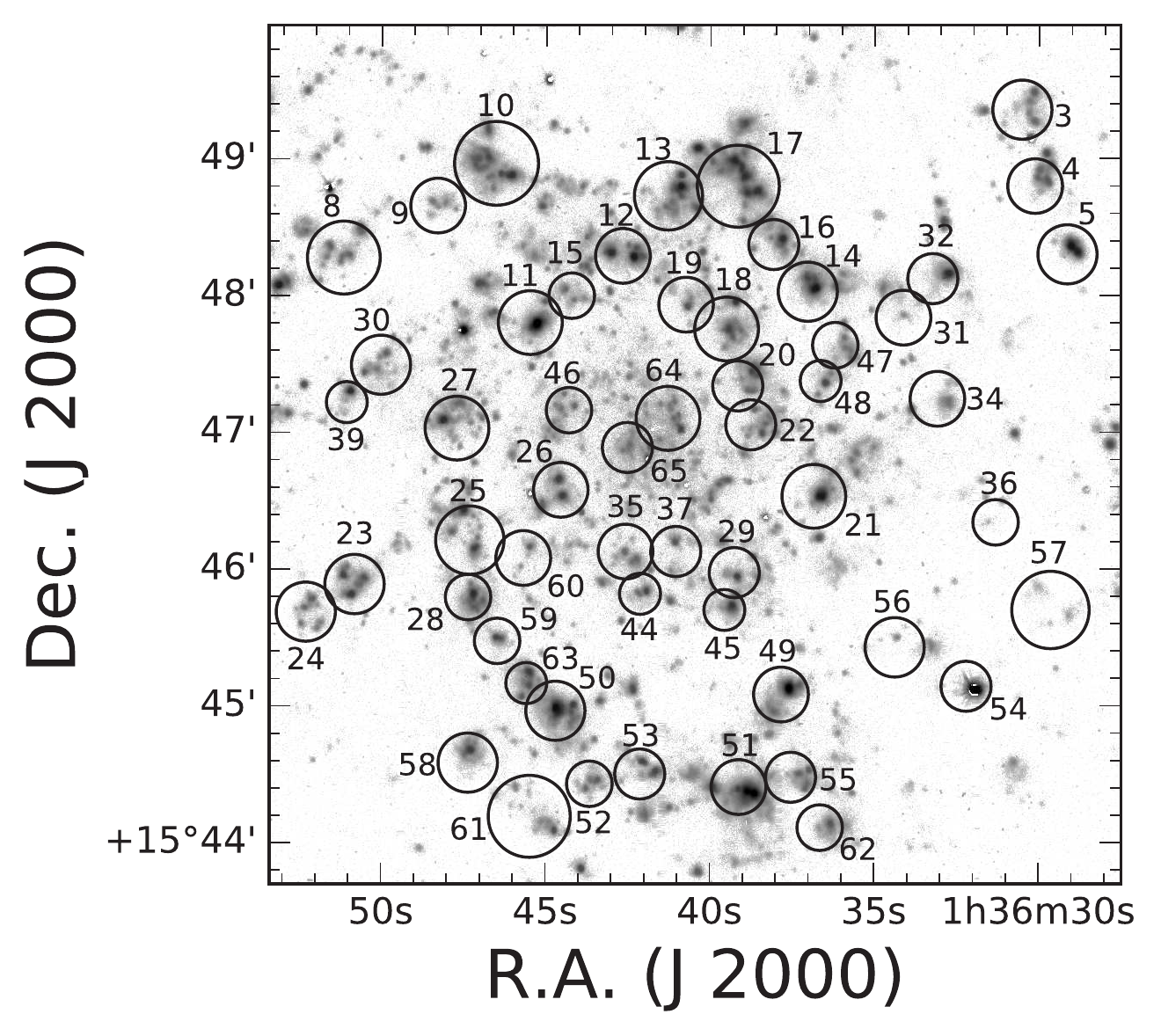}
\includegraphics[width=0.3\textwidth]{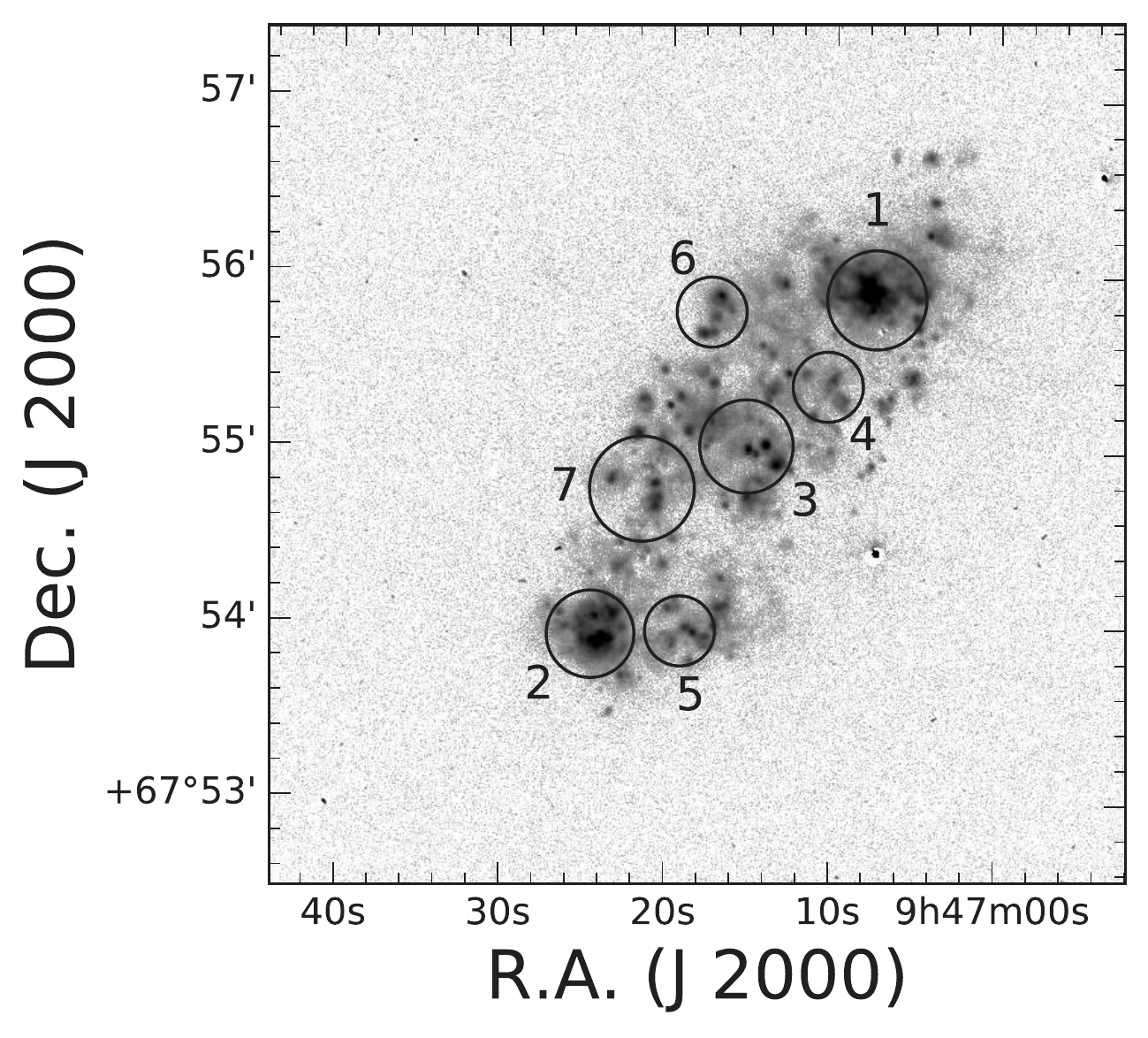}
\includegraphics[width=0.33\textwidth]{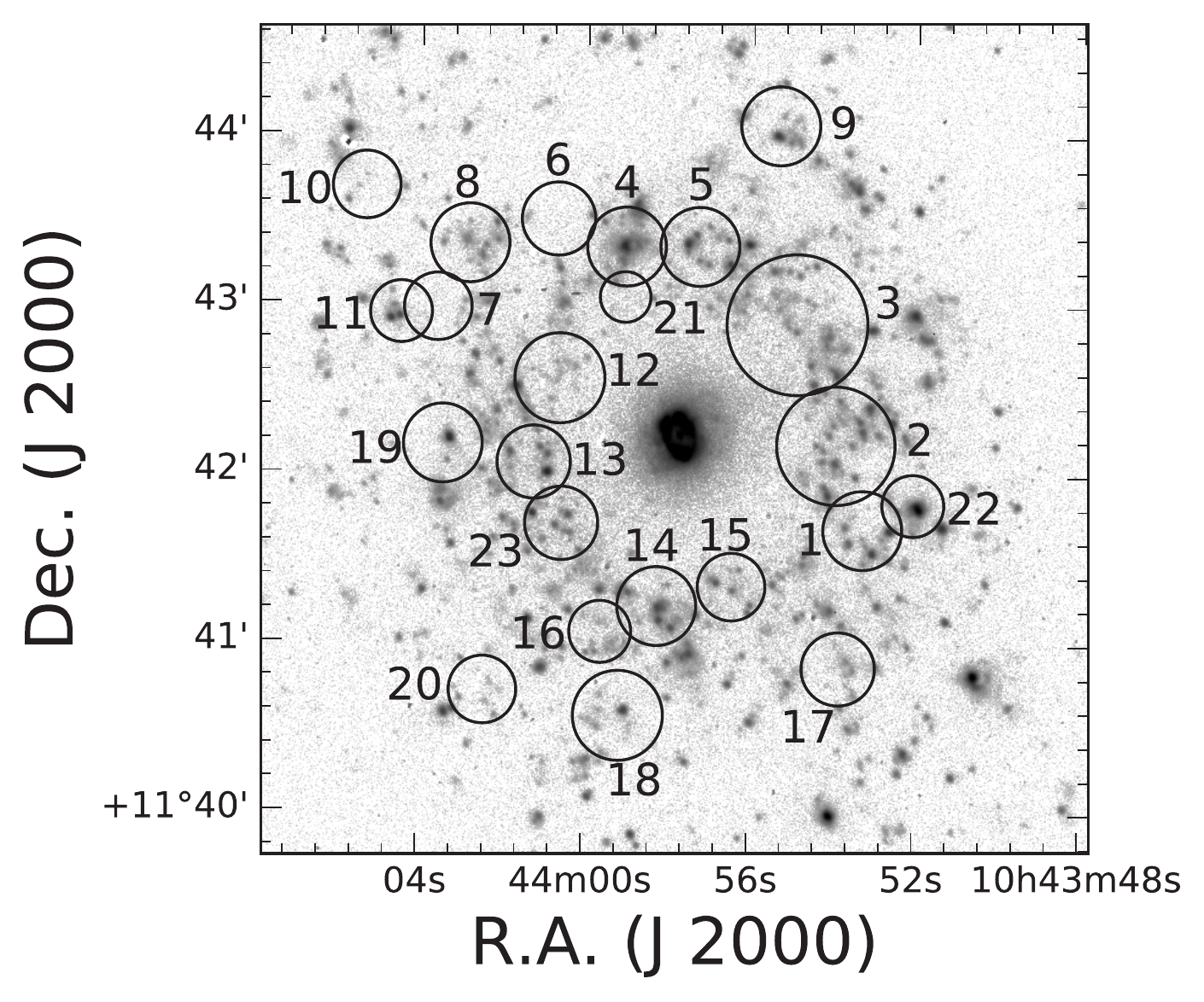}
\caption{H$\alpha$ images of NGC~628, NGC~2976, and NGC~3351 with the apertures superposed.
\hfill}
\label{maps}
\end{figure*}

Since we used data from different telescopes with different angular resolutions, we reduced all the data to the resolution in the far-IR (160 $\mu$m) in \cite{ourpreviouswork}, using the convolution procedure and kernels presented in \cite{2011Aniano}. Since we used some results from \cite{ourpreviouswork} in our current study, we also reduced the H$\alpha$ data to this same resolution (12"). In \cite{ourpreviouswork}, the regions for which we performed aperture photometry were chosen such that emission in a specific region would be observed in at least one of the ranges considered. The numbers of SFCs studied were 65 in NGC~628, 7 in NGC~2976, and 23 in NGC~3351. The apertures used for each galaxy are shown in \ref{maps}. The observations at 8 and 24 $\mu$m were used to determine the stellar background that was subtracted. As in\cite{ourpreviouswork}, results are presented here for the corrected fluxes at 8 and 24 $\mu$m. The aperture areas were calculated using distance estimates of 9.77 Mpc for NGC~628, 3.63 Mpc for NGC~2976, and 10.57 Mpc for NGC~3351 \cite{2016AJ....152...50T}.

The data cubes in the CO line were taken from the HERACLES survey. A spectrum was constructed for every SFC, and the standard deviation $\sigma$ was calculated for sections free of CO lines. We took the velocity spread $\Delta V$ in a particular SFC to be the difference of the velocities at the end points of the spectrum where the line intensity exceeded $3~\sigma$. If the spectrum of a SFC had no intensities exceeding $3~\sigma$, the object was excluded from consideration. We also discarded SFCs where $\Delta V$ exceeded half the global velocity variation in the given galaxy. In this point, our study differs from \cite{2016Mogotsi}, where the velocity dispersion was determined by fitting Gaussians to the spectra. We considered not only complexes with pronounced single- or two-peaked CO emission, but also those with weak emission, where it was not always possible to distinguish one or two peaks. Therefore, we employed the simpler procedure described above. Obviously, from a mathematical point of view, $\Delta V$ is not a dispersion, and we accordingly will call this quantity the ``velocity scatter''.

Table 1 summarizes the observational parameters of the SFCs derived in the present study.

\section{RESULTS OF THE APERTURE PHOTOMETRY}
One of our main aims was to analyze the relationships (or lack there of) between the H$\alpha$ data and various parameters of the emission of the SFCs in these three galaxies, including parameters that can be used as indicators of the SFR. Therefore, we will first consider the correlations obtained using this data. In the plots below, the SFCs in different galaxies are indicated by pale orange (NGC~628), blue (NGC~2976), and green (NGC~3351) symbols.

\subsection{H$\alpha$ Emission of the SFCs}
\begin{figure*}
\includegraphics[scale=0.3]{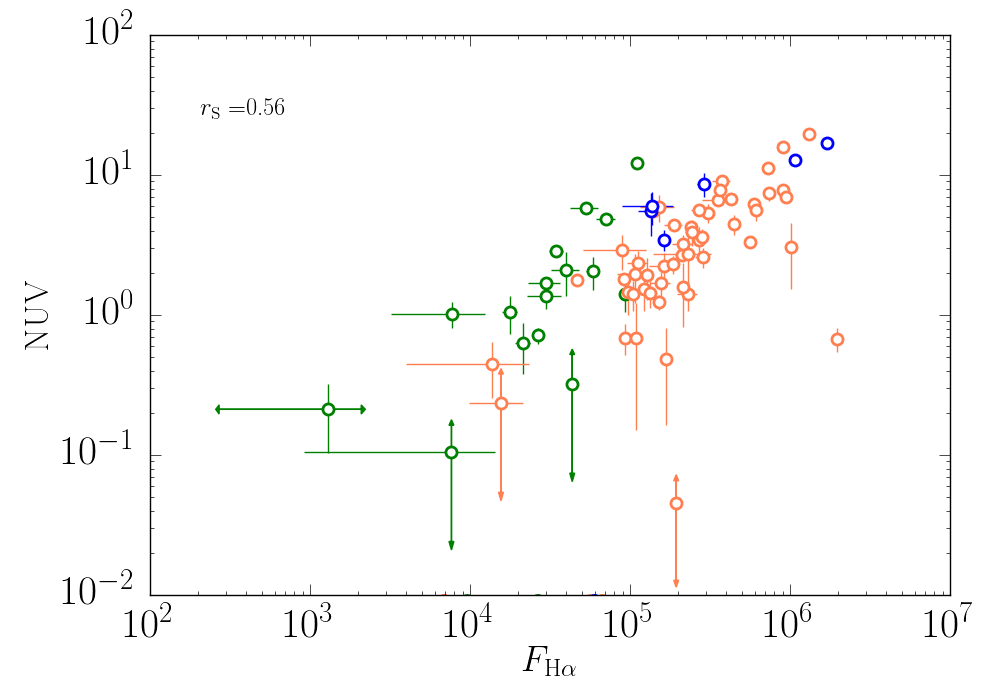}
\includegraphics[scale=0.3]{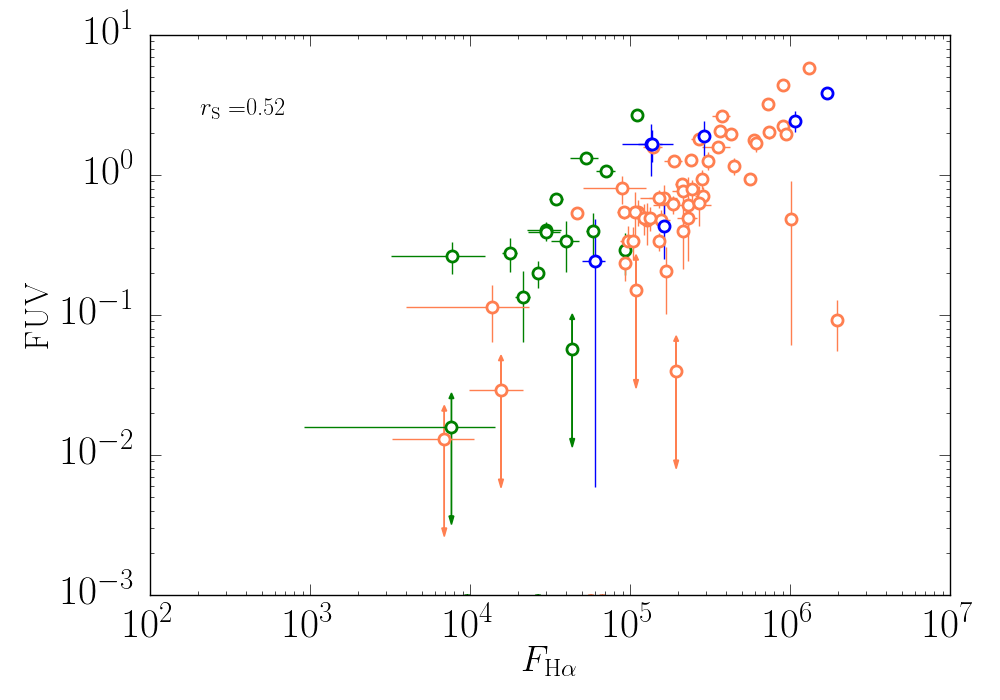}
\includegraphics[scale=0.3]{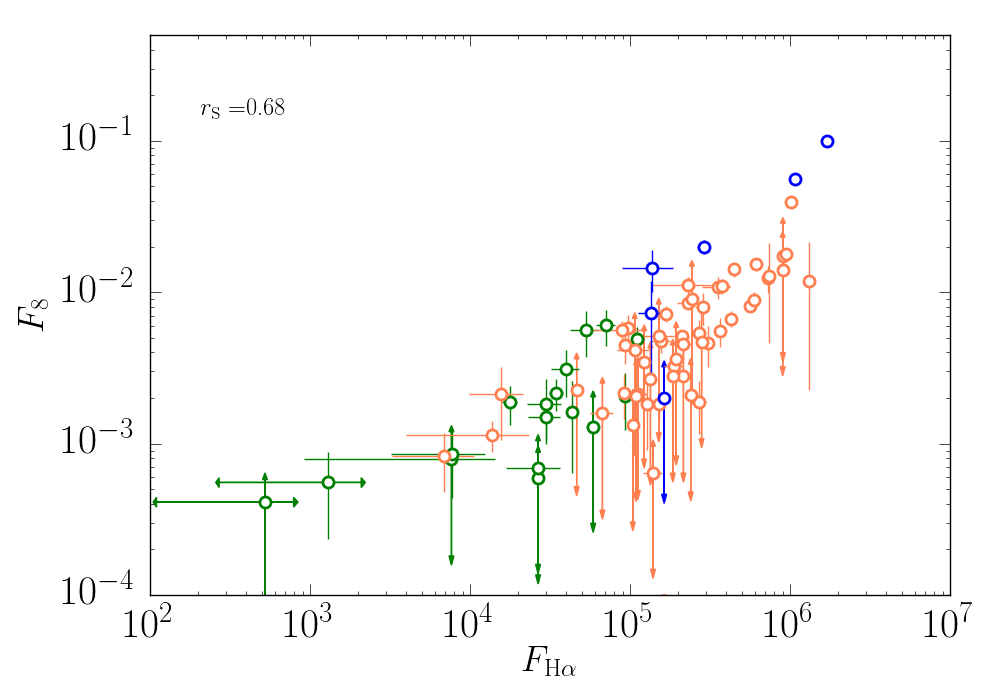}
\includegraphics[scale=0.3]{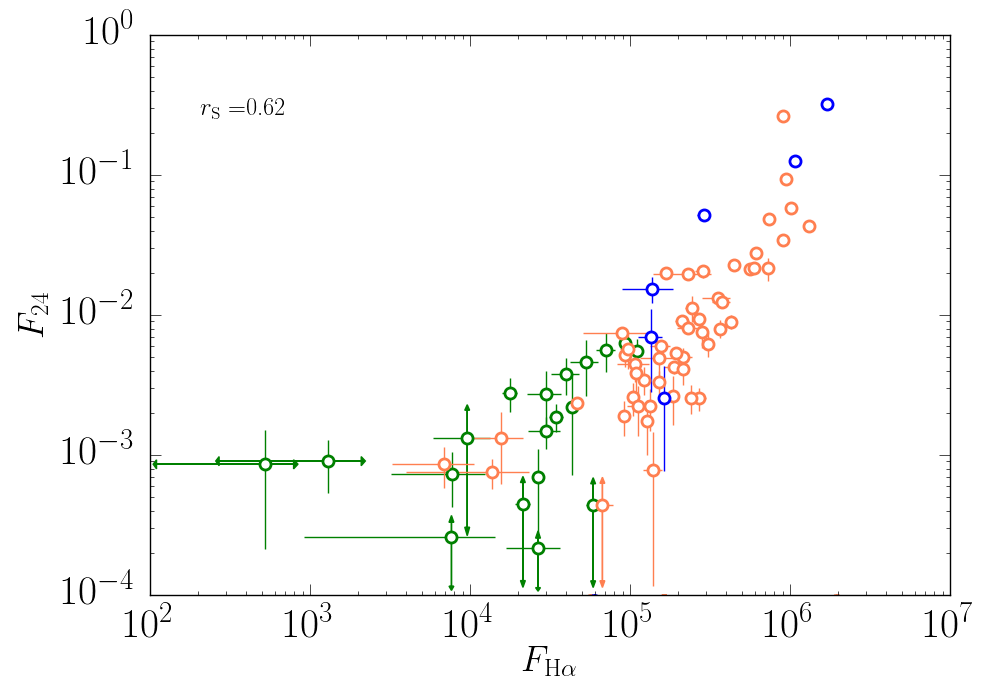}
\includegraphics[scale=0.3]{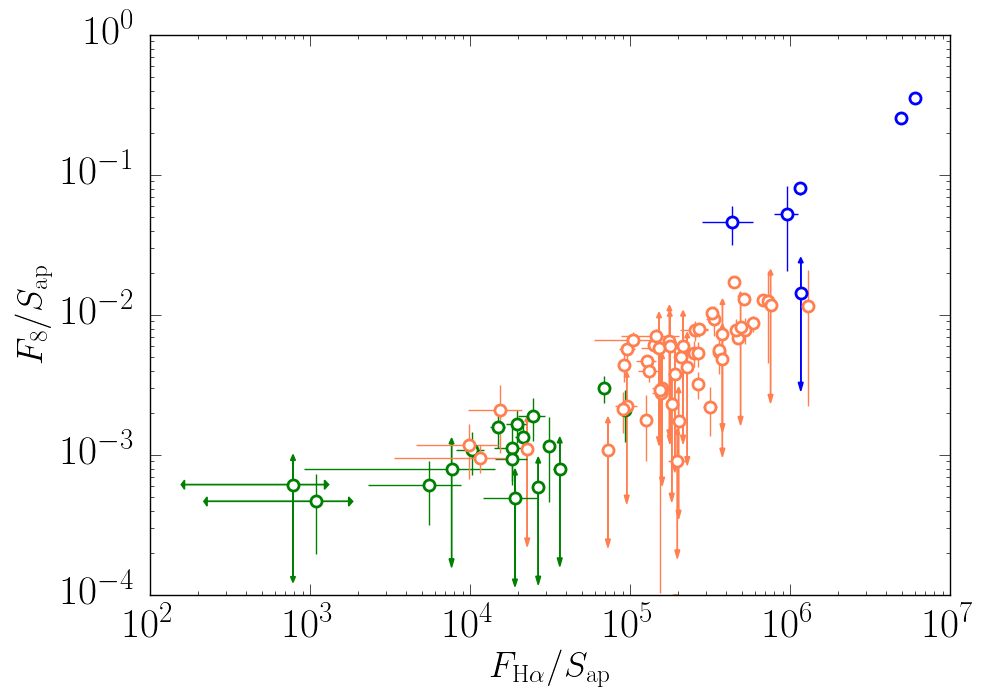}
\includegraphics[scale=0.3]{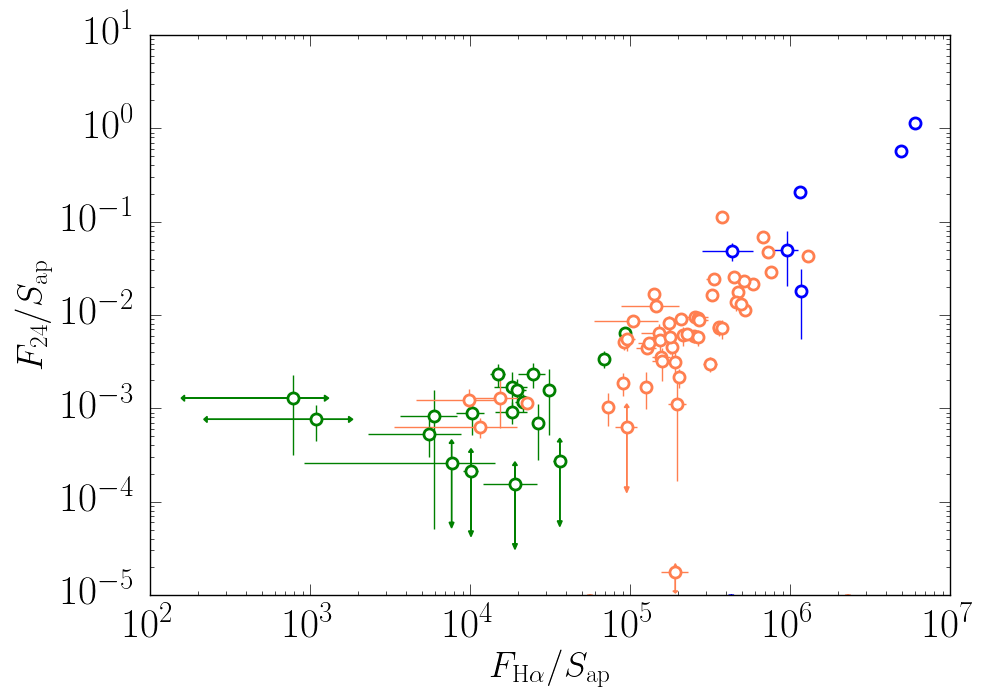}
\caption{Relationship between the fluxes in the H$\alpha$ line and the fluxes in the GALEX NUV (top left) and the FUV (top right) filters, and in the Spitzer 8 $\mu$m (middle left) and 24 $\mu$m (middle right) bands. The lower panels are similar to the middle plots, but plot surface brightnesses instead of fluxes. The SFCs in NGC~628, NGC~2976, and NGC~3351 are plotted in pale orange, blue, and green respectively.}
\label{Fig1}
\end{figure*}

The upper panels of Fig.~\ref{Fig1} show the relationship between the fluxes in the H$\alpha$ line and in the GALEX NUV and FUV filters. Here and below, the H$\alpha$ flux is presented in arbitrary units. It is obvious that the UV and H$\alpha$ emission are well correlated. Correlations are also observed between the H$\alpha$ flux and the Spitzer 8 $\mu$m and 24 $\mu$m fluxes (Fig.~\ref{Fig1}, middle panels). The emission in these bands is probably related to PAHs and UV-absorbing dust particles near zones of ionized hydrogen. It is therefore not surprising that the H$\alpha$ emission correlates well with both the UV and mid-IR emission. Note that the dynamic range covered by the available data is quite wide, about three orders of magnitude in all the photometric bands considered. The fluxes of the SFCs in NGC 3351 are usually an order of magnitude lower than the corresponding mean values in the other two galaxies.

SFCs 20 and 21 in NGC~3351, which have the lowest H$\alpha$ fluxes, fall away from the general trend in the $F_8-F_{{\rm H}\alpha}$ and $F_{24}-F_{{\rm H}\alpha}$ diagrams. Their special position is particularly noticeable in the bottom panels of Fig.~\ref{Fig1}, where we have plotted the surface brightnesses rather than the fluxes; i.e., the fluxes divided by the aperture areas (in $kpc^{2}$). The surface brightness of these two complexes at 24 $\mu$m significantly exceeds the expected value for the low amount of H$\alpha$ emission observed in them. Moreover, their spatial locations in the galaxy does not stand out in any way (see Fig.~\ref{maps}).

\begin{figure*}
\includegraphics[scale=0.3]{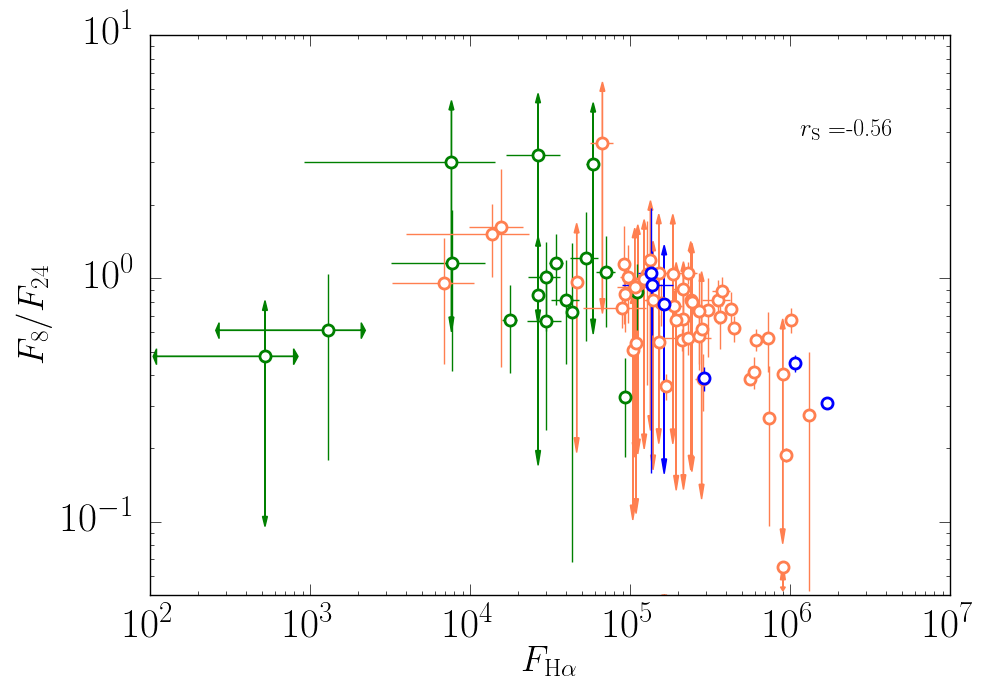}
\includegraphics[scale=0.3]{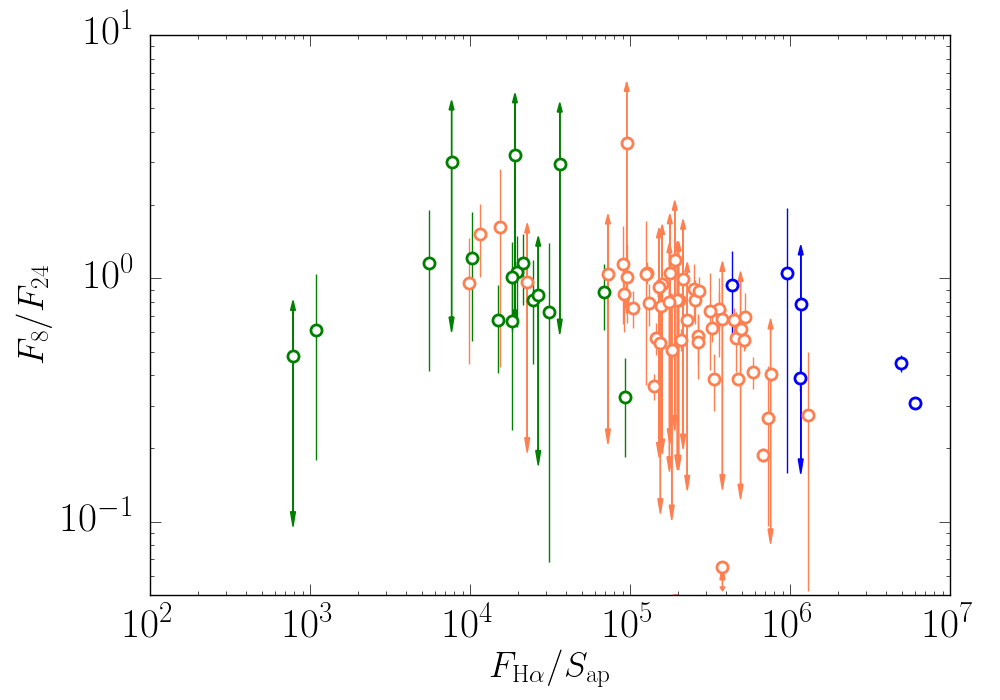}
\includegraphics[scale=0.3]{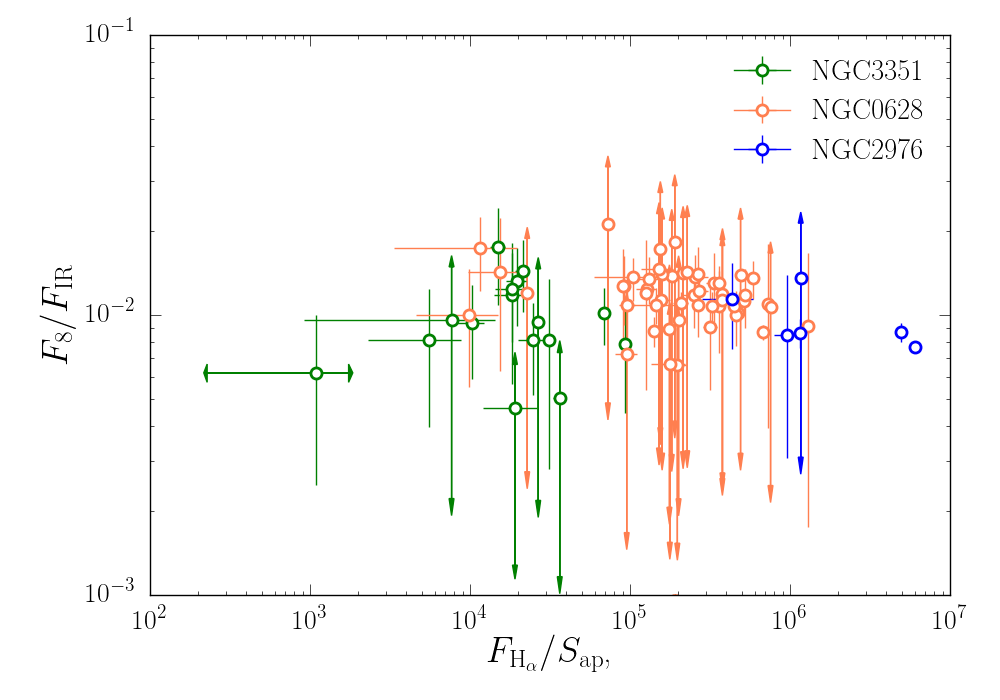}
\includegraphics[scale=0.3]{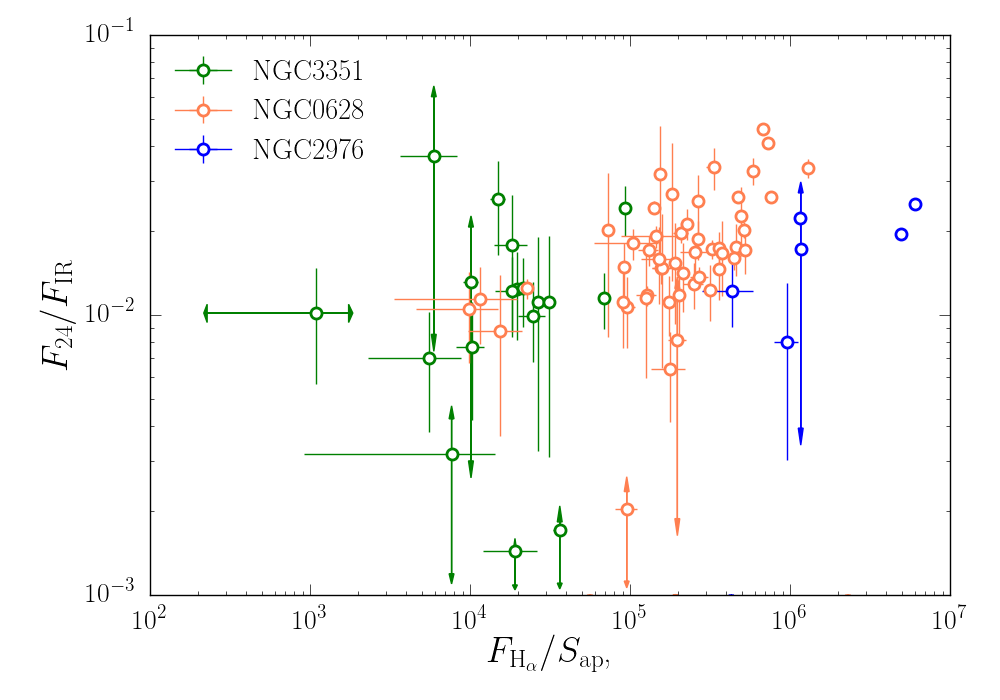}
\caption{Upper: ratio of the fluxes at 8 and 24 $\mu$m as a function of the H$\alpha$ flux (left) and the H$\alpha$ surface brightness (right). Lower: dependences of the ratio of the 8 $\mu$m flux and far-IR flux (left) and of the ratio of the 8 $\mu$m flux and far-IR flux (right) on the H$\alpha$ surface brightness.}
\label{Fig2}
\end{figure*}

The data in \ref{Fig1} show only that the fluxes of the SFCs in different ranges are mutually correlated. A comparison of the flux ratios may be more informative here. \ref{Fig2} shows how the IR flux ratio $F_{8}/F_{24}$ correlates with the H$\alpha$ flux (upper left panel) and with the H$\alpha$ surface brightness (upper right panel). As was shown in \cite{2013MNRAS.431.2006K}, the $F_{8}/F_{24}$ $\mu$m flux ratio can be used as an indicator of the relative PAH abundance. The upper left panel of Fig.~\ref{Fig2} shows an anti-correlation between $F_{8}/F_{24}$ and the H$\alpha$ flux. This is quite expected if a higher intensity of UV radiation brings about more effective destruction of PAHs. A similar anti-correlation between $F_{8}/F_{24}$ and the H$\alpha$ surface brightness can be seen in the upper right panel of Fig.~\ref{Fig2}.

The reasons for this anti-correlation are clarified in the lower panels of Fig.~\ref{Fig2}, where the F8/FIR and $F_{24}/F_{\rm IR}$ flux ratios are plotted against the H$\alpha$ surface brightness, where $F_{\rm IR}= F_{70}+F_{100}+F_{160}$ is the total flux in the far-IR. The $F_8/F_{\rm IR}$ ratio is essentially independent of the H$\alpha$ surface brightness. The situation concerning the $F_{24}/F_{\rm IR}$ ratio is more complex. If we consider the entire set of SFCs, there is a weak correlation between this ratio and the H$\alpha$ surface brightness: the brighter the H$\alpha$ emission, the higher the relative contribution of the mid-IR to the total IR emission. However, if we consider the SFCs in NGC 3351 separate, this could lead to a different conclusion. In general, the anti-correlation between $F_{8}/F_{24}$ and the H$\alpha$ emission is more dependent on the behavior of the IR emission at 24 $\mu$m than on the emission at 8 $\mu$m.

The same two complexes in NGC 3351 stand out in the upper panels of Fig.~\ref{Fig2} as in Fig.~\ref{Fig1}. The collected data indicate that these SFCs are abnormally bright in the 24 $\mu$m band for their low H$\alpha$ fluxes GALEX UV fluxes. Neither the procedure used to isolate these complexes, nor their locations in the galaxy are unusual. Thus, identifying the reasons for their difference from the rest of the SFCs in our sample requires additional study.

\begin{figure*}
\includegraphics[scale=0.3]{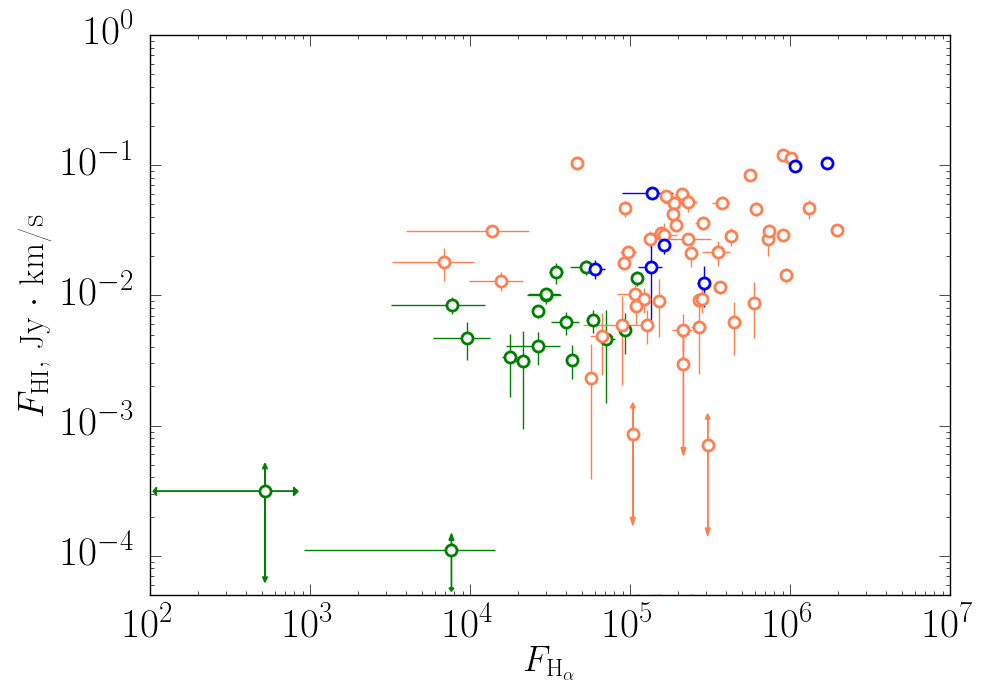}
\includegraphics[scale=0.3]{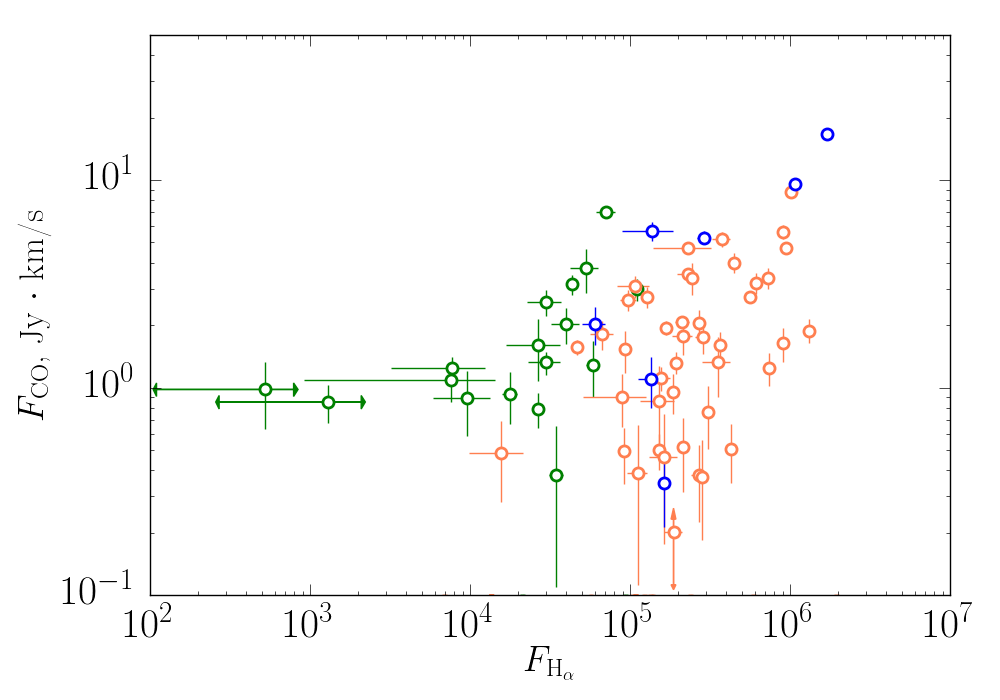}
\includegraphics[scale=0.3]{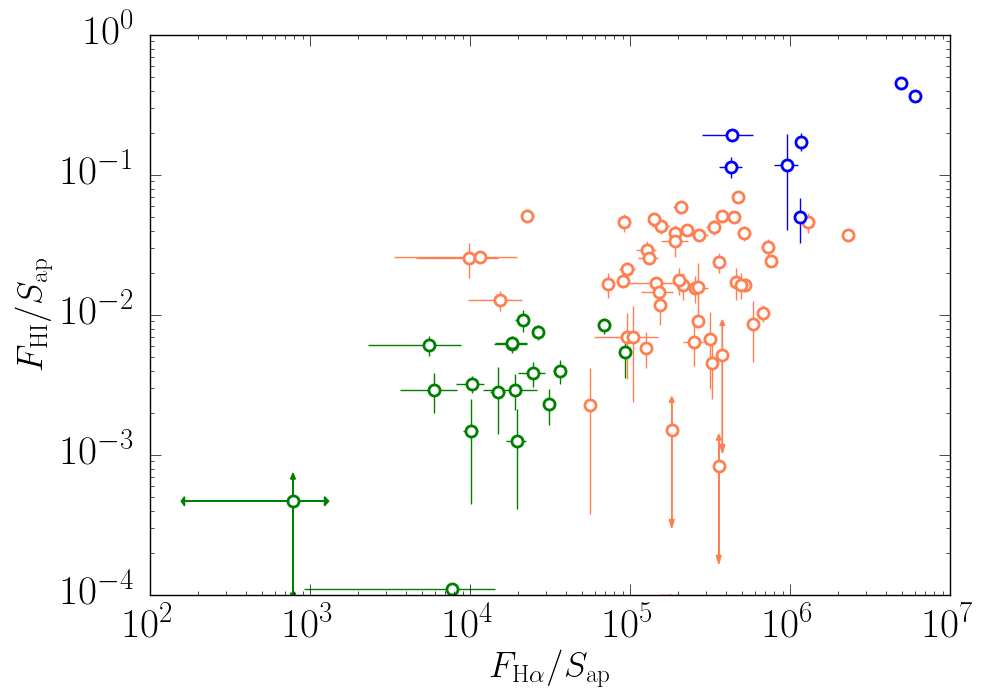}
\includegraphics[scale=0.3]{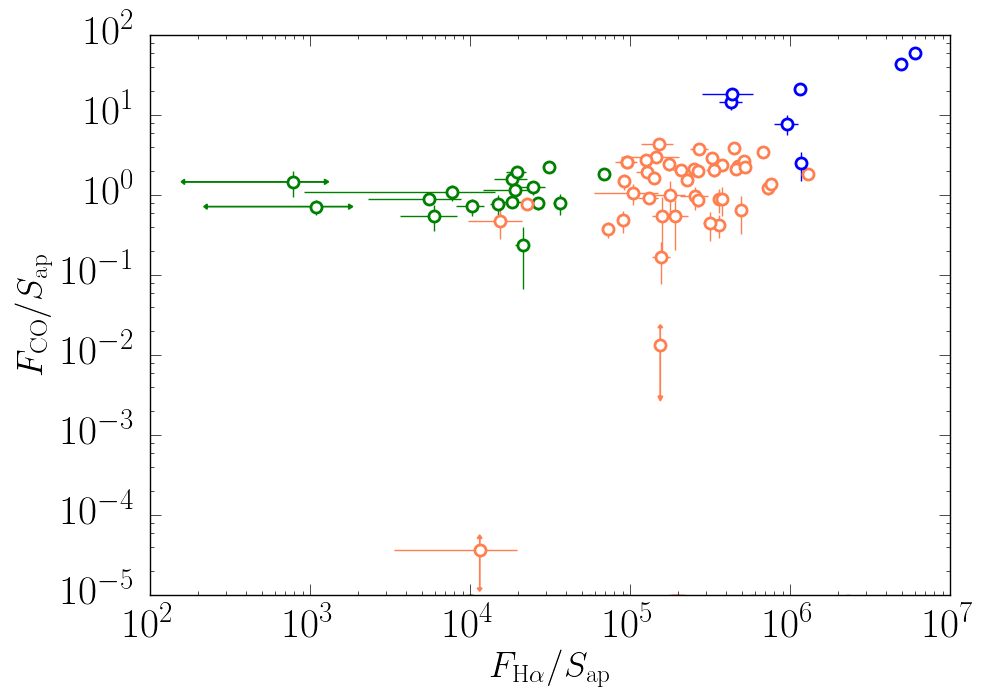}
\caption{Relationship between the fluxes in the HI line (top left) and the CO line (upper right) and the H$\alpha$ flux; the lower panels are analogous, but plot the corresponding surface brightnesses.}
\label{Fig3}
\end{figure*}

Figure~\ref{Fig3} shows relationships between indicators of the presence of interstellar gas: the fluxes in the H$\alpha$, HI, and CO lines. The upper panels plot the fluxes and the lower panels the corresponding surface brightnesses. Two SFCs in NGC~628 (56 and 59) with abnormally low abundances of molecular gas were excluded From the plots for the CO data. The SFCs in NGC~3351 stand out from the main distribution of points in all four diagrams. Although complexes with low emission in the H$\alpha$ and HI lines stand out somewhat from the set of all the SFCs, on average, they follow the general trend. The situation is different for two complexes that stand out in the diagrams for the CO line, which prove to be the same SFCs with low H$\alpha$ emission noted earlier; at the same time, their CO fluxes and surface brightnesses are the same as those in the other SFCs.

\subsection{Kinematics of CO}
The best known property of the kinematics of interstellar gas is the so-called Larson relation between the velocity dispersion (the width of the molecular lines) and the spatial scale \cite{1981MNRAS.194..809L}. In our case, the spatial scales considered are very large, close to the largest anticipated dimensions of the SFCs (see,e.g., \cite{2014MNRAS.442.3711G}). Therefore, our sample contains both SFCs in which CO emission is observed in a single line and those in which there are several distinct CO lines. Therefore, as was noted already, we are referring to $\Delta V$ as a velocity scatter, not the velocity dispersion. Furthermore, by virtue of the lack of a formal procedure, we cannot accurately assess the uncertainty in $\Delta V$ . However, we can assume that this uncertainty is comparable to the $\Delta V$ values for SFCs with low FCO fluxes and for SFRs for which $\Delta V$ is comparable to the total velocity range considered.

\begin{figure*}
\includegraphics[scale=0.6]{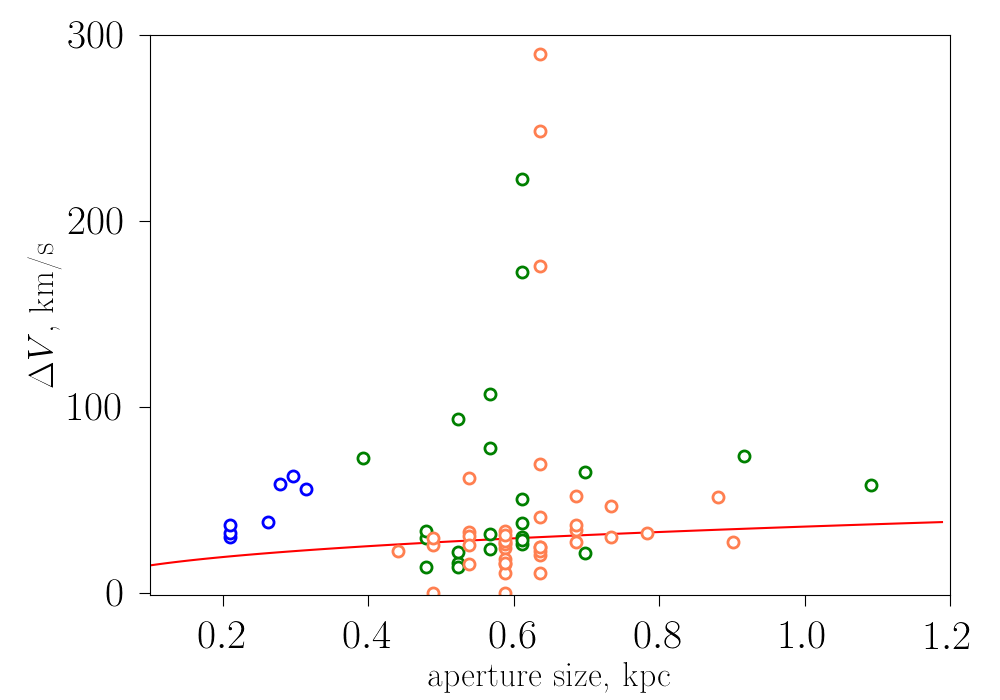}
\caption{Relationship between the velocity scatter and the aperture size.}
\label{dvaper}
\end{figure*}

\begin{figure*}
\includegraphics[scale=0.6]{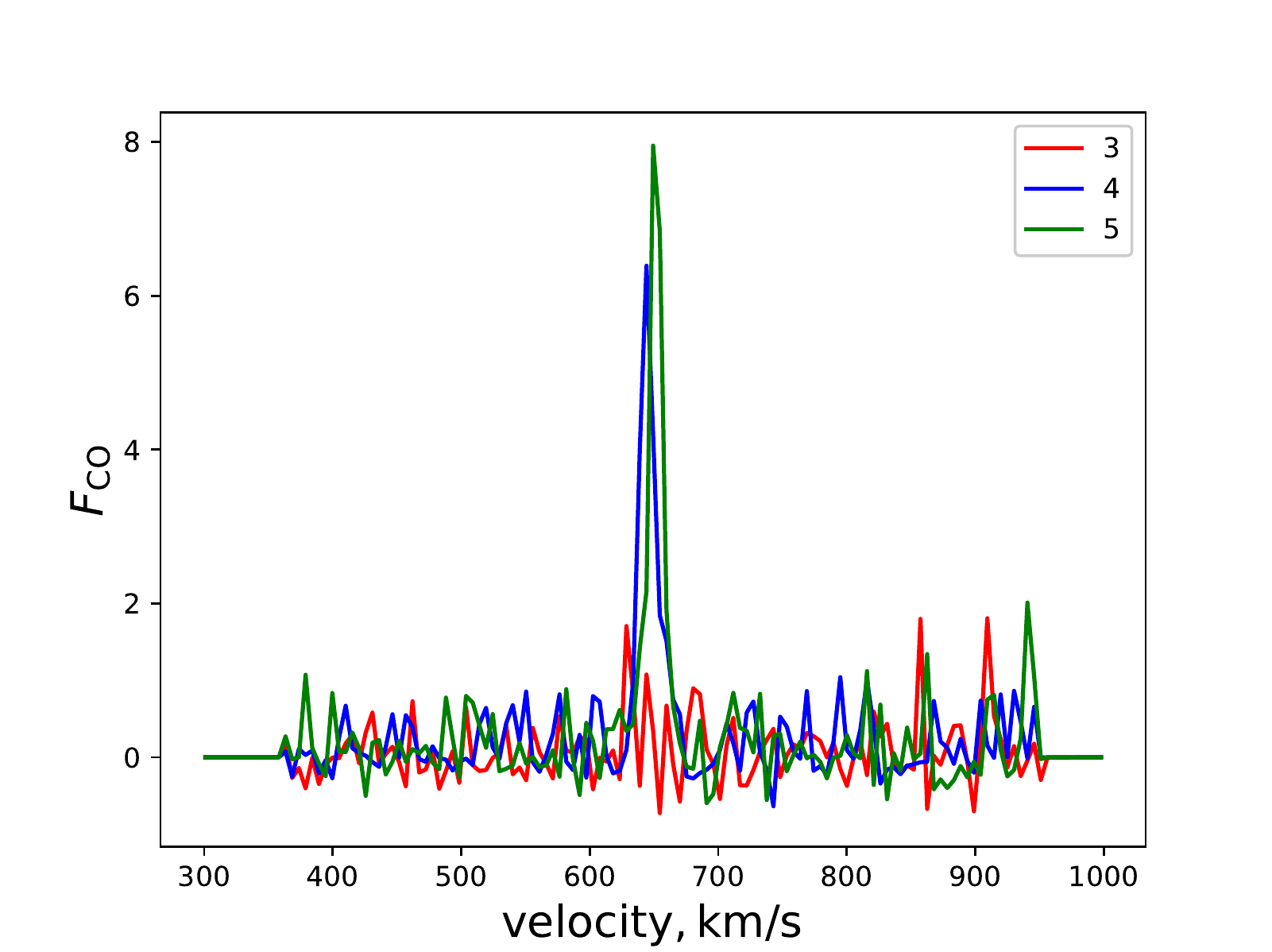}
\caption{Profiles of CO line in SFCs 3, 4, and 5 in NGC~628.}
\label{prof}
\end{figure*}

Figure~\ref{dvaper} shows the relationship between dimensions of a SFC and the velocity scatter in it. The straight line shows the Larson relation from \cite{1981MNRAS.194..809L}. In our sample, this ratio appears to describe the lower envelope of the obtained points. In some of the SFCs, $\Delta V$ significantly exceeds the predictions of the ``classical'' relation, namely complexes 3, 6, and 7 in NGC~628 and complexes 8 and 19 in NGC~3351. All these complexes are located at the peripheries of these galaxies. However, their locations and their high $\Delta V$ values are unlikely to be related, since the velocity scatters in other peripheral SFCs are significantly lower. As an example, Fig.~\ref{prof} shows CO line profiles for areas $3–5$ of NGC~628. All three complexes are located in the same region at the periphery of the galaxy, but the CO emission profiles in them are quite different. In the complexes 4 and 5, a single intense line with $\Delta V =$ 24-25~km/s is observed, whereas the CO emission of complex 3 has several lines slightly above the background and separated by a velocity interval of almost 300 km/s.

\begin{figure*}
\includegraphics[scale=0.6]{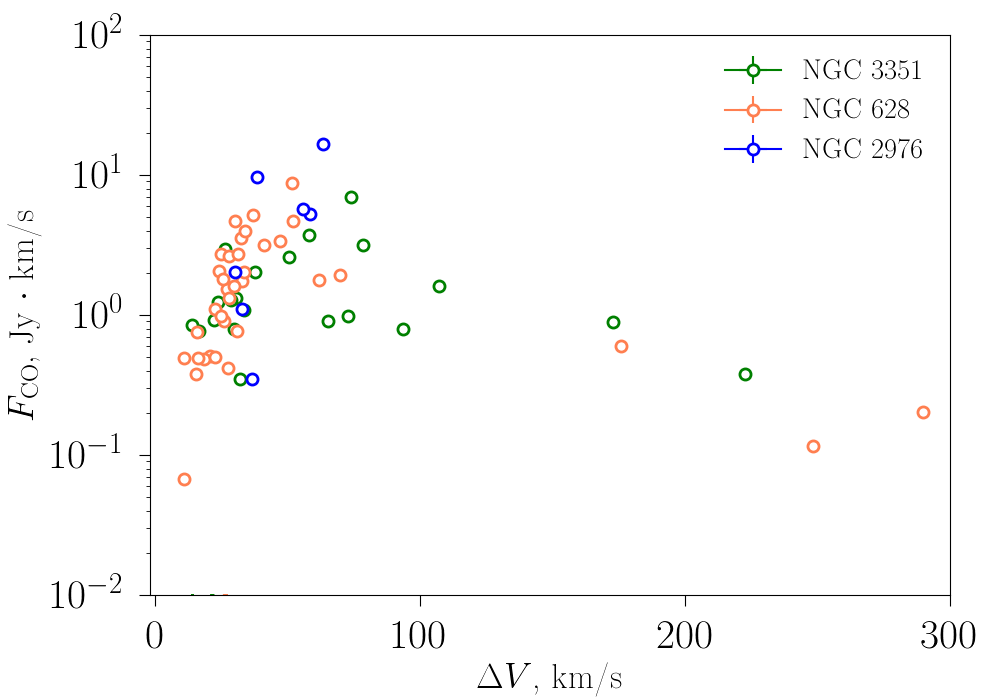}
\caption{Relationship between the flux in the CO line and the velocity scatter.}
\label{dvco}
\end{figure*}

Figure~\ref{dvco} shows the relationship between the velocity scatter $\Delta V$ and the flux in the CO line. We can distinguish two groups of SFCs in this figure. In objects with $\Delta V\lesssim70$ km/s, the CO flux increases with increasing $\Delta V$ , whereas, in objects with higher velocity scatters, the CO flux drops with increasing $\Delta V$ . It may be that, in the first group, we are dealing with molecular gas that remains largely undisturbed by star formation. In this case, we would expect a growth of $F_{\rm CO}$ with increasing $\Delta V$ , since the CO flux can be considered a measure of the mass of the SFC. The decrease in the CO flux with increasing $\Delta V$ in the second group may reflect the circumstance that it includes SFCs whose gas is already appreciably disturbed by star-formation processes.

\begin{figure}
\includegraphics[width=0.7\textwidth]{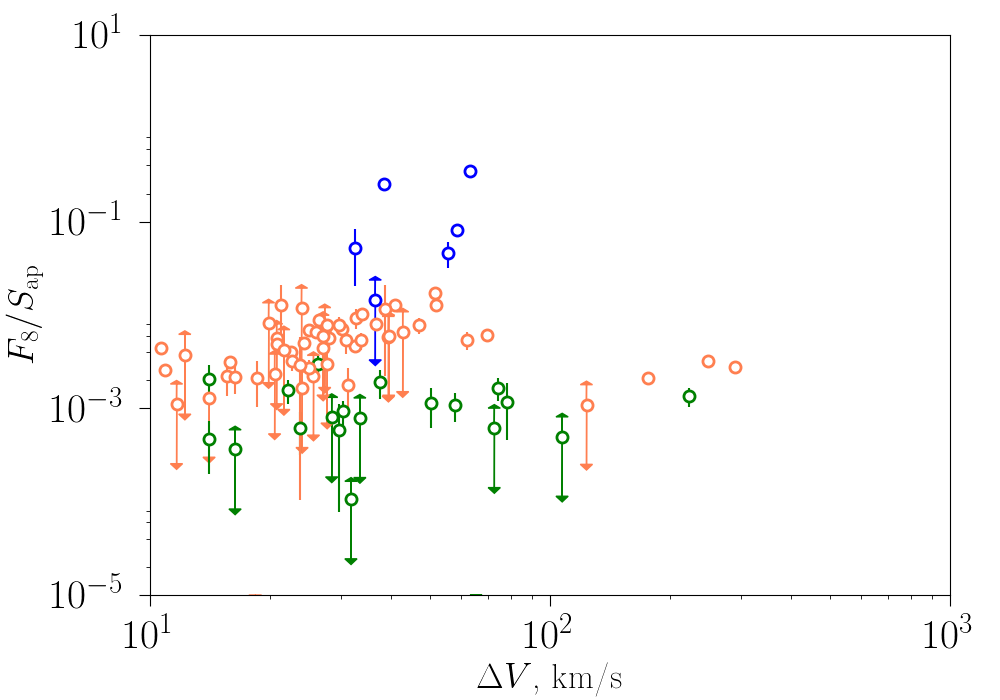}
\includegraphics[width=0.7\textwidth]{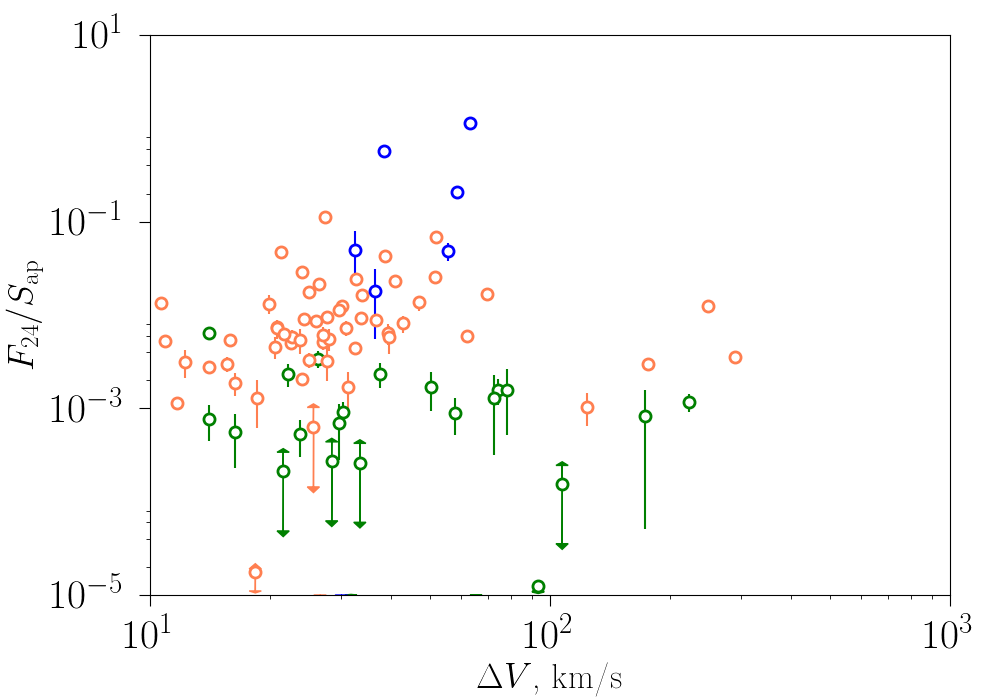}
\includegraphics[width=0.7\textwidth]{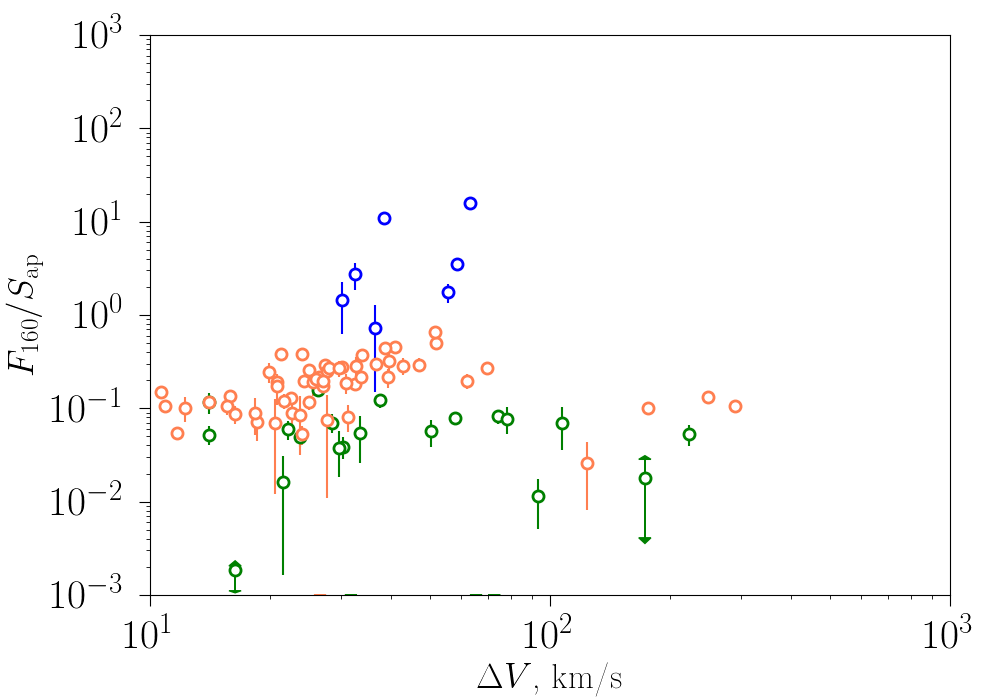}
\caption{Relationship between the surface brightnesses in the IR and the velocity scatter.\hfill}
\label{otherfluxes}
\end{figure}

Similar behavior is also demonstrated by the IR fluxes. Figure~\ref{otherfluxes} shows how the surface brightnesses at 8, 24, and 160 $\mu$m are related to the velocity scatter. Here, also, two groups of SFCs are seen: in the first group, with $\Delta V\lesssim70$ km/s, the surface brightness increases with the velocity scatter; in the second group, with higher values of $\Delta V$ , the velocity scatter is large, while the surface brightnesses are small.

\begin{figure*}
\includegraphics[scale=0.6]{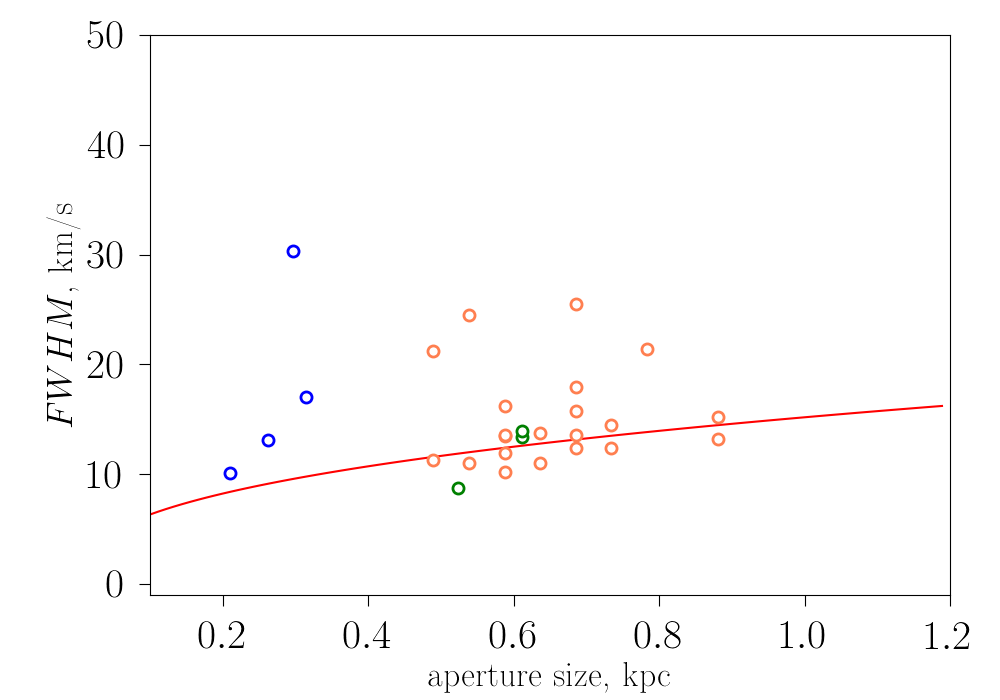}
\caption{Relationship between the velocity scatter and the aperture size.}
\label{fwhmaper}
\end{figure*}

As an additional test, we selected areas where the CO emission is characterized by a single peak and fitted this peak with a Gaussian in order to find the full width at half-maximum of the line, FWHM (or the velocity scatter). Comparison of the estimated line widths and the velocity dispersions we derived previously showed that, in the case of single-peaked lines, the FWHM and $\Delta V$ values agree well. In Fig.~\ref{fwhmaper}, the estimated line width is compared to the SFC aperture size. The Larson relation is shown by the line in this figure, and agrees well with the parameters of the SFCs.

%

\section{DISCUSSION}
Currently, one of the main directions in studies of the ISMis detailed simulations of Galactic and extragalactic molecular clouds and regions of star formation, including galaxies at high redshifts, taking into account feedback effects (see, e.g.,  \cite{2018PASJ...70S..56L,2018MNRAS.477.2716K,2018ApJ...859...68K}). This enables exploration of the evolution of the molecular gas and star-formation processes under a wide range of conditions, in different environments, in galaxies with different morphological types, etc. Observationally, the different relations between the spatial scales, the kinematic characteristics, and other properties of the studied objects become criteria for the adequacy of the theoretical models. For us, the initial motivation for such studies was also the selection of factors that can have a significant impact on the evolution of the ensemble of dust particles in star-forming regions. We expect that, in the initial stages of the evolution of an SFC, one such factor will be the UV radiation of massive stars, whereas, in later stages, a more important factor will be the destruction of dust particles by shocks from supernova explosions.

In order to test these hypotheses, and also to obtain overall estimates of the interrelations between the various parameters of SFCs, we attempted to find relationships between the fluxes of SFCs in various spectral ranges. Here, we considered not only the fluxes themselves, whose strength could be associated with the masses of the corresponding SFCs, but also the surface brightnesses, i.e., the flux per physical area of the aperture in $pc^{2}$. Since, in fact, a SFC is a three-dimensional structure, its surface brightness could also be related to its mass (albe it to a lesser extent than the flux), due to the extent of the complex along the line of sight.

The radiation fluxes, which are in one way or another associated with the presence of massive hot stars, correlate well with each other. In particular, a correlation is observed between the H$\alpha$ flux and the GALEX FUV and NUV fluxes, as well as between the corresponding surface brightnesses. Because the emission at 8 and 24 $\mu$m is presumably due to macromolecules and very fine dust particles that are heated by the absorption of individual UV photons, the radiation in these ranges also correlates well with the fluxes and surface brightnesses in H$\alpha$ and the FUV and NUV bands.

Although the fluxes at 8 $\mu$m and at 24 $\mu$m are both correlated with the H$\alpha$ flux, the nature of these correlations is different, as is underscored by the anticorrelation between the H$\alpha$ flux and the $F_{8}/F_{24}$ flux ratio. To obtain a tentative identification of the origins of this behavior, we applied the results of calculations made in \cite{2007Draine}. In this model, one characteristic of the radiation field that causes heating of dust particles is the parameter $U_{\min}$, representing the average radiation intensity in the SFC in units of the radiation field in the solar neighborhood. In the absence of evolutionary effects, that is, if the characteristics of the ensemble of dust particles are constant in time, the $F_{8}/F_{24}$ flux ratio is essentially independent of $U_{\min}$(for the limiting values of this parameter we have considered). In addition, increasing $U_{\min}$ in the model \cite{2007Draine} leads to a slight increase in $F_{8}/F_{\rm IR}$ and $F_{24}/F_{\rm IR}$.

This is obviously not what we see in observations (the variable  $U_{\min}$, defined in  \cite{ourpreviouswork}, correlates well with $F_{{\rm H}\alpha}$, making these variables interchangeable). First,
the ratio $F_{8}/F_{24}$ drops substantially with increasing
$F_{{\rm H}\alpha}$ and $U_{\min}$. Second, with increasing $F_{{\rm H}\alpha}$ and $U_{\min}$, the ratio $F_{24}/F_{\rm IR}$ increases substantially more rapidly than is predicted by the model  \cite{2007Draine}. The absence of an observed dependence of $F_{8}/F_{\rm IR}$ on $F_{{\rm H}\alpha}$  does not contradict the weak increase in this quantity with increasing $U_{\min}$ in the model.

We can attempt to explain these discrepancies by supplementing the model with evolutionary effects. For example, the decrease in $F_{8}/F_{24}$ with increasing radiation intensity could be explained by destruction of the PAH macromolecules. Indeed, in the model \cite{2007Draine}, the observed decrease in the ratio $F_{8}/F_{24}$ can be explained by a decrease in the mass fraction of PAHs $q_{\rm PAH}$ from 4.6\% to 0.5\%. However, such a decrease in qPAH should result in an even more significant decline of  $F_{8}/F_{\rm IR}$, which is not observed.

Another explanation could be associated with the other parameter of the model \cite{2007Draine}, the mass fraction $\gamma$ of the dust that is irradiated with intensities higher than $U_{\min}$. In fact, $\gamma$ determines the fraction of the dust that is in the direct vicinity of the stars that are the sources of the ionizing radiation. The model \cite{2007Draine} predicts that increasing $\gamma$ with $U_{\min}$ fixed leads to a reduction of the ratio $F_{8}/F_{24}$, and to a significant increase in $F_{24}/F_{\rm IR}$ with $F_{8}/F_{\rm IR}$ remaining almost constant. This corresponds to the observations, but this correspondence should not be overemphasized. The values of $U_{\min}$ and $\gamma$ derived from the observations in \cite{ourpreviouswork} are obtained by fitting spectra, and are therefore functions of the observed IR fluxes.

Our estimated values of $\Delta V$ appreciably exceed the velocity dispersions obtained in other similar studies. For example, the maximum CO velocity dispersions in  \cite{2016Mogotsi} do not exceed 40 km/s, and similar values for the ionized gas are obtained in \cite{2015MNRASMoiseev}. CO line widths up to 50 km/s are noted in \cite{1998ApJFrail}. However, this is due to the fact that these studies involved objects that are either relatively bright sources of CO or H$\alpha$ emission, or have significantly smaller spatial scales (as in \cite{1998ApJFrail}). The method of distinguishing SFCs
using several spectral ranges makes it possible to identify complexes that are not bright sources of CO or H$\alpha$ emission, and simultaneously display a significant internal velocity scatter, exceeding 100 km/s.

Such values seem quite plausible. Of course, such velocities (and also velocities of the order of 50 km/s) would not be expected in the molecular gas of SFCs, if the only factor for feedback is zones of ionized hydrogen. However, such velocities could be encountered if supernova explosions have already begun in a complex. Though there are no direct observations of such velocities, there are some indirect indications of their existence. For example, the observed destruction of dust particles during interactions between supernovae remnants and molecular clouds \cite{2010ApJArendt,2006ApJTappe,2012ApJTappe} suggests the presence of shocks with velocities exceeding 100 km/s \cite{2010A&AMicelotta,2016ARepMurga}. Velocities exceeding 100 km/s are also obtained in numerical calculations of the expansion of supernovae remnants in a cloudy environment (see, e.g., \cite{2017ApJSlavin}).

There is no question that the parameter $\Delta V$ we have used is determined significantly more poorly than a velocity dispersion obtained by fitting a Gaussian to an observed line profile. This uncertainty is amplified by the fact that, when the CO emission in a SFC can no longer be characterized by a single profile, the intensity of this emissions weakens and the identification of individual peaks becomes more difficult. However, this enhances interest in the gas kinematics in this stage of the SFC evolution, when the feedback effects disrupting the original molecular structure of the complex apparently become important. We believe that our results are interesting in this respect, as they emphasize the need for further research in this direction, although they do not yet enable us to draw firm conclusions. They indicate that the relationship between $\Delta V$ and other parameters of the ISM and star-forming indicators can be complex, and that investigation of these relationships requires
studies of a more extensive sample of galaxies and SFCs and analysis of other lines, primarily, HI and H$\alpha$. A preliminary interpretation of such data may provide evidence that, in addition to the obvious correlation between $\Delta V$ and the CO luminosity, whose basis may be the mass of the object (see, e.g., \cite{2017A&A...599A..76M}), there also exists another regime in which the increase of $\Delta V$ is related to decreasing luminosity, which may have an evolutionary nature. These regimes are worthy of a separate study.

\section{CONCLUSION}
The main conclusions of the study can be summarized as follows.
\begin{enumerate}
\item We have found correlations between the H$\alpha$ fluxes and surface brightnesses of SFCs and analogous parameters in the UV and IR.
\item The ratio of the fluxes of SFCs in the 8 and 24 $\mu$m bands decreases with increasing H$\alpha$ flux. This is most likely due to a more substantial increase of the 24 $\mu$m flux.
\item Our analysis of the origins of the variations $F_{8}/F_{24}$ flux ratio indicates that this may be due not only to evolutionary factors (variations in the mass fraction of PAHs), but also to variations in the conditions for excitation of this emission.
\item Our analysis of data on the kinematics of molecular gas in the SFCs shows that the relationship between the CO luminosity and the velocity scatter is ambiguous. Preliminary results indicate that the CO luminosity and $\Delta V$ are correlated when $\Delta V\lesssim70$~km/s, while the growth of $\Delta V$ is accompanied by a decrease in the CO luminosity for larger values of the velocity scatter. However, these results require further verification due to uncertainties associated with estimation of the parameter $\Delta V$ .
\end {enumerate}

\section{FUNDING}
This work has been supported by Program 211 of the Government of the Russian Federation (agreement No. 02.A03.21.0006), the Ministry of Science of the Russian Federation (main part of the Government contract No. AAAA-A17-117030310283-7), and Basic Research Program No. 12 of the Russian Academy of Sciences, ``The Origin and Evolution of the Universe''.

\section{ACKNOWLEDGMENTS}
We thank the referee for important comments and useful improvements of the quality of the manuscript. This study was based in part on observational data obtained by the Spitzer Space Telescope, which is operated by the Jet Propulsion Laboratory, California Institute of Technology, under a contract with NASA. Herschel is an ESA space observatory with science instruments provided by European-led Principal Investigator consortia and with important participation from NASA. The National Radio Astronomy Observatory is a facility of the National Science Foundation operated under cooperative agreement by Associated Universities, Inc. This work is also based on observations carried out with the IRAM NOEMA Interferometer [30m telescope]. IRAM is supported by INSU/CNRS (France), MPG (Germany) and IGN (Spain).


\begin{thebibliography}{99}
\bibitem{1998ARA&A..36..189K} Kennicutt, R.~C.\ 1998, \araa, 36, 189.
\bibitem{2000A&A...358..869R} Rocha-Pinto, H.~J., Scalo, J., Maciel, W.~J., et al.\ 2000, \aap, 358, 869.
\bibitem{2010ApJ...710L..11R} Robitaille, T.~P., \& Whitney, B.~A.\ 2010, \apjl, 710, L11.
\bibitem{2013seg..book..419C} Calzetti, D.\ 2013, Secular Evolution of Galaxies, 419.
\bibitem{2014AstL...40..278W} Wiebe, D.~S., Khramtsova, M.~S., Egorov, O.~V., et al.\ 2014, Astronomy Letters, 40, 278.
\bibitem{2014MNRAS.444..757K} M.~S.~Khramtsova, D.~S.~Wiebe, T.~A.~Lozinskaya, and O.~V.~Egorov, 2014, Monthly Not. Roy. Astron. Soc., 444, 757.
\bibitem{2017Bendo} Bendo, G.~J., Miura, R.~E., Espada, D., et al.\ 2017, \mnras, 472, 1239.
\bibitem{2018Audcent-Ross} Audcent-Ross, F.~M., Meurer, G.~R., Wong, O.~I., et al.\ 2018, \mnras, 480, 119.
\bibitem{2010Calzetti} Calzetti, D., Wu, S.-Y., Hong, S., et al.\ 2010, \apj, 714, 1256.
\bibitem{THINGS} F.~Walter et al., 2008, Astron. J., 136, 2563. 
\bibitem{kingfish} R.~C.~Kennicutt et al., 2011, Publ. Astron. Soc. Pacif., 123, 1347.
\bibitem{sings} R.~C.~Kennicutt et al., 2003, Publ. Astron. Soc. Pacif., 115, 928. 
\bibitem{heracles} A.~K.~Leroy et al., 2009, Astron. J., 137, 4670.
\bibitem{ourpreviouswork} Smirnova, K.~I., Murga, M.~S., Wiebe, D.~S., et al.\ 2017, Astronomy Reports, 61, 646.
\bibitem{2013Caldu-Primo} Cald{\'u}-Primo, A., Schruba, A., Walter, F., et al.\ 2013, \aj, 146, 150.
\bibitem{2013Pety} Pety, J., Schinnerer, E., Leroy, A.~K., et al.\ 2013, \apj, 779, 43.
\bibitem{2016Mogotsi} Mogotsi, K.~M., de Blok, W.~J.~G., Cald{\'u}-Primo, A., et al.\ 2016, \aj, 151, 15.
\bibitem{Swartz2006} Swartz, D.~A., Yukita, M., Tennant, A.~F., et al.\ 2006, \apj, 647, 1030.
\bibitem{Elmegreen1997} Elmegreen, D.~M., Chromey, F.~R., Santos, M., et al.\ 1997, \aj, 114, 1850.
\bibitem{2014Sanchez-Blazques} S{\'a}nchez-Bl{\'a}zquez, P., Rosales-Ortega, F., Diaz, A., et al.\ 2014, \mnras, 437, 1534.
\bibitem{2013Gusev} Gusev, A.~S., \& Efremov, Y.~N.\ 2013, \mnras, 434, 313.
\bibitem{2008ApJS..178..247K} Kennicutt, R.~C., Lee, J.~C., Funes, J.~G., et al.\ 2008, \apjs, 178, 247.
\bibitem{2005ApJ...619L...1M} Martin, D.~C., Fanson, J., Schiminovich, D., et al.\ 2005, \apjl, 619, L1.
\bibitem{2011Aniano} G.~Aniano, B.~T.~Draine, K.~D.~Gordon, and K.~Sandstrom, 2011, Publ. Astron. Soc. Pacif., 123, 1218.
\bibitem{2016AJ....152...50T} Tully, R.~B., Courtois, H.~M., \& Sorce, J.~G.\ 2016, \aj, 152, 50.
\bibitem{2013MNRAS.431.2006K} M.~S.~Khramtsova, D.~S.~Wiebe, P.~A.~Boley, and Y.~N.~Pavlyuchenkov, 2013, Monthly Not. Roy. Astron. Soc., 431, 2006.
\bibitem{1981MNRAS.194..809L} Larson, R.~B.\ 1981, \mnras, 194, 809.
\bibitem{2014MNRAS.442.3711G} Gusev, A.~S.\ 2014, \mnras, 442, 3711.
\bibitem{2018PASJ...70S..56L} Li, Q., Tan, J.~C., Christie, D., et al.\ 2018, \pasj, 70, S56.
\bibitem{2018MNRAS.477.2716K} Krumholz, M.~R., Burkhart, B., Forbes, J.~C., et al.\ 2018, \mnras, 477, 2716.
\bibitem{2018ApJ...859...68K} Kim, J.-G., Kim, W.-T., \& Ostriker, E.~C.\ 2018, \apj, 859, 68.
\bibitem{2007Draine} B.~T.~Draine and A.~Li, 2007, Astrophys. J., 657, 810.
\bibitem{2015MNRASMoiseev} Moiseev, A.~V., Tikhonov, A.~V., \& Klypin, A.\ 2015, \mnras, 449, 3568
\bibitem{1998ApJFrail} Frail, D.~A., \& Mitchell, G.~F.\ 1998, \apj, 508, 690.
\bibitem{2010ApJArendt} Arendt, R.~G., Dwek, E., Blair, W.~P., et al.\ 2010, \apj, 725, 585.
\bibitem{2006ApJTappe} Tappe, A., Rho, J., \& Reach, W.~T.\ 2006, \apj, 653, 267.
\bibitem{2012ApJTappe} Tappe, A., Rho, J., Boersma, C., et al.\ 2012, \apj, 754, 132.
\bibitem{2010A&AMicelotta} Micelotta, E.~R., Jones, A.~P., \& Tielens, A.~G.~G.~M.\ 2010, \aap, 510, A36.
\bibitem{2016ARepMurga} Murga, M.~S., Khoperskov, S.~A., \& Wiebe, D.~S.\ 2016, Astronomy Reports, 60, 669.
\bibitem{2017ApJSlavin} Slavin, J.~D., Smith, R.~K., Foster, A., et al.\ 2017, \apj, 846, 77.
\bibitem{2017A&A...599A..76M} Melnick, J., Telles, E., Bordalo, V., et al.\ 2017, \aap, 599, A76.

\end{thebibliography}

\clearpage

\newpage

\tiny
\begin{center}
\begin{longtable}{|c|c|c|c|c|c|c|c|} 
\caption[Parameters of the interstellar medium of galaxies: NGC~628, NGC~2976, NGC~3351.]{Parameters of the ISM of galaxies: NGC~628, NGC~2976, NGC~3351.} \label{par_HoII} \\ 

\hline \multicolumn{1}{|c|}{\textbf{$n$}} & \multicolumn{1}{c|}{\textbf{$\alpha$}, h : m : s} & \multicolumn{1}{c|}{\textbf{$\delta$},$^{\circ}$ : $'$ : $''$} & \multicolumn{1}{c|}{\textbf{$F_{Ha}	\pm	\Delta F_{Ha}$}} & \multicolumn{1}{c|}{\textbf{$\Delta V$}} & \multicolumn{1}{c|}{\textbf{$FUV\pm\Delta FUV$}, Jy}  & \multicolumn{1}{c|}{\textbf{$NUV\pm\Delta NUV$}, Jy}  \\ \hline 
\endfirsthead 

\multicolumn{7}{c}%
{\tablename\ \thetable\ -- parameters of galaxies: NGC628,NGC2976, NGC3351} \\ 
\hline \multicolumn{1}{|c|}{\textbf{$n$}} & \multicolumn{1}{c|}{\textbf{$\alpha$}} & \multicolumn{1}{c|}{\textbf{$\delta$}} & \multicolumn{1}{c|}{\textbf{$F_{Ha}	\pm	\Delta F_{Ha}$}} & \multicolumn{1}{c|}{\textbf{$\Delta V$}} & \multicolumn{1}{c|}{\textbf{$FUV\pm\Delta FUV$}}  & \multicolumn{1}{c|}{\textbf{$NUV\pm\Delta NUV$}}  \\ \hline 
\endhead 
\hline 
\endfoot 
\multicolumn{7}{|c|}{NGC 628} \\
\hline

1	&	1:36:37.090	&	+15:50:26.34	&				&	15.893	&	0.90	$\pm$	0.07	&	2.74	$\pm$	0.24	\\
2	&	1:36:35.856	&	+15:50:06.39	&				&	10.683	&	1.44	$\pm$	0.07	&	4.86	$\pm$	0.22	\\
3	&	1:36:30.524	&	+15:49:21.48	&	187112	$\pm$	23370.1	&	289.815	&	1.26	$\pm$	0.07	&	4.36	$\pm$	0.14	\\
4	&	1:36:30.129	&	+15:48:47.99	&	212226	$\pm$	22471.5	&	24.200	&	0.86	$\pm$	0.06	&	2.70	$\pm$	0.20	\\
5	&	1:36:29.143	&	+15:48:18.06	&	564871	$\pm$	3554.1	&	24.951	&	0.93	$\pm$	0.02	&	3.30	$\pm$	0.09	\\
6	&	1:36:41.632	&	+15:49:59.98	&				&	175.873	&	1.36	$\pm$	0.05	&	4.21	$\pm$	0.17	\\
7	&	1:36:53.233	&	+15:48:00.97	&				&	248.354	&	0.94	$\pm$	0.03	&	3.25	$\pm$	0.10	\\
8	&	1:36:51.160	&	+15:48:16.65	&	229183	$\pm$	33775.9	&	32.474	&	0.49	$\pm$	0.10	&	1.40	$\pm$	0.34	\\
9	&	1:36:48.297	&	+15:48:39.46	&	93207.6	$\pm$	3807.32	&	27.101	&	0.24	$\pm$	0.06	&	0.69	$\pm$	0.17	\\
10	&	1:36:46.520	&	+15:48:57.99	&	903213	$\pm$	31982.2	&	27.346	&	2.22	$\pm$	0.22	&	7.83	$\pm$	0.63	\\
11	&	1:36:45.483	&	+15:47:48.16	&	938270	$\pm$	12135.7	&	51.948	&	1.96	$\pm$	0.07	&	7.01	$\pm$	0.26	\\
12	&	1:36:42.669	&	+15:48:17.38	&	593233	$\pm$	28153.5	&	26.454	&	1.77	$\pm$	0.14	&	6.19	$\pm$	0.46	\\
13	&	1:36:41.287	&	+15:48:43.74	&	724125	$\pm$	49259.1	&	47.089	&	3.21	$\pm$	0.18	&	11.25	$\pm$	0.66	\\
14	&	1:36:37.041	&	+15:48:01.70	&	617392	$\pm$	38329.7	&	41.028	&	1.70	$\pm$	0.23	&	5.62	$\pm$	0.90	\\
15	&	1:36:44.224	&	+15:47:59.92	&	67248.5	$\pm$	10703.4	&	25.627	&	0.00	$\pm$	0.08	&	0.00	$\pm$	0.32	\\
16	&	1:36:38.078	&	+15:48:22.36	&	284264	$\pm$	28259.7	&	32.821	&	0.71	$\pm$	0.12	&	2.60	$\pm$	0.42	\\
17	&	1:36:39.164	&	+15:48:48.01	&	1015670	$\pm$	54566.8	&	51.666	&	0.48	$\pm$	0.42	&	3.04	$\pm$	1.50	\\
18	&	1:36:39.510	&	+15:47:45.31	&	353517	$\pm$	69948.6	&	27.670	&	1.58	$\pm$	0.13	&	6.60	$\pm$	0.50	\\
19	&	1:36:40.744	&	+15:47:56.00	&	97590.5	$\pm$	10707.1	&	27.944	&	0.34	$\pm$	0.10	&	1.45	$\pm$	0.45	\\
20	&	1:36:39.164	&	+15:47:20.37	&	306121	$\pm$	21634.3	&	30.848	&	1.27	$\pm$	0.18	&	5.37	$\pm$	0.81	\\
21	&	1:36:36.845	&	+15:46:31.92	&	448142	$\pm$	29824	&	33.891	&	1.16	$\pm$	0.16	&	4.45	$\pm$	0.70	\\
22	&	1:36:38.769	&	+15:47:03.27	&	213806	$\pm$	29916.6	&	61.956	&	0.77	$\pm$	0.14	&	3.23	$\pm$	0.50	\\
23	&	1:36:50.813	&	+15:45:53.44	&	428126	$\pm$	9013.18	&	20.740	&	1.95	$\pm$	0.09	&	6.70	$\pm$	0.24	\\
24	&	1:36:52.293	&	+15:45:41.32	&	169082	$\pm$	5249.07	&	69.655	&	0.20	$\pm$	0.10	&	0.48	$\pm$	0.32	\\
25	&	1:36:47.308	&	+15:46:12.68	&	229383	$\pm$	89808.9	&	30.223	&	0.61	$\pm$	0.36	&	2.71	$\pm$	1.13	\\
26	&	1:36:44.544	&	+15:46:34.78	&	271693	$\pm$	21488.5	&	33.623	&	0.64	$\pm$	0.20	&	3.41	$\pm$	0.86	\\
27	&	1:36:47.704	&	+15:47:01.85	&	375431	$\pm$	48296.6	&	36.767	&	2.64	$\pm$	0.19	&	9.04	$\pm$	0.80	\\
28	&	1:36:47.358	&	+15:45:47.75	&	367558	$\pm$	29952.1	&	29.737	&	2.07	$\pm$	0.11	&	7.83	$\pm$	0.50	\\
29	&	1:36:39.263	&	+15:45:58.44	&	88877.4	$\pm$	37840.1	&	25.981	&	0.81	$\pm$	0.18	&	2.91	$\pm$	0.80	\\
30	&	1:36:50.024	&	+15:47:29.63	&	156049	$\pm$	22769	&	22.509	&	0.48	$\pm$	0.09	&	1.68	$\pm$	0.35	\\
31	&	1:36:34.129	&	+15:47:50.29	&	15625.6	$\pm$	5818.48	&	18.563	&	0.03	$\pm$	0.18	&	0.24	$\pm$	0.63	\\
32	&	1:36:33.240	&	+15:48:07.39	&	270672	$\pm$	28531.3	&	15.537	&	1.80	$\pm$	0.15	&	5.60	$\pm$	0.54	\\
33	&	1:36:27.959	&	+15:46:55.41	&				&		&	0.59	$\pm$	0.06	&	1.68	$\pm$	0.25	\\
34	&	1:36:33.093	&	+15:47:14.67	&	91981.8	$\pm$	7698.48	&	16.291	&	0.54	$\pm$	0.05	&	1.81	$\pm$	0.18	\\
35	&	1:36:42.570	&	+15:46:07.70	&	127808	$\pm$	10047.9	&	31.339	&	0.48	$\pm$	0.16	&	1.92	$\pm$	0.63	\\
36	&	1:36:31.316	&	+15:46:20.51	&	6907.14	$\pm$	3646.19	&		&	0.01	$\pm$	0.03	&	0.00	$\pm$	0.10	\\
37	&	1:36:41.040	&	+15:46:07.69	&	72394.6	$\pm$	6417.32	&	29.241	&	0.00	$\pm$	0.12	&	0.00	$\pm$	0.44	\\
38	&	1:36:53.183	&	+15:47:04.68	&				&	24.966	&	0.33	$\pm$	0.07	&	1.23	$\pm$	0.27	\\
39	&	1:36:51.061	&	+15:47:13.24	&	151484	$\pm$	9221.33	&	22.640	&	0.34	$\pm$	0.05	&	1.24	$\pm$	0.16	\\
40	&	1:36:55.749	&	+15:46:18.36	&				&	10.930	&	0.97	$\pm$	0.16	&	2.85	$\pm$	0.50	\\
41	&	1:36:54.120	&	+15:46:07.68	&				&	24.011	&	0.61	$\pm$	0.03	&	1.94	$\pm$	0.08	\\
42	&	1:36:59.698	&	+15:46:16.92	&				&		&	2.87	$\pm$	0.04	&	8.89	$\pm$	0.21	\\
43	&	1:36:57.823	&	+15:46:59.68	&				&	14.058	&	5.67	$\pm$	0.13	&	18.81	$\pm$	0.39	\\
44	&	1:36:42.126	&	+15:45:49.18	&	104133	$\pm$	5565.47	&	20.495	&	0.34	$\pm$	0.09	&	1.41	$\pm$	0.34	\\
45	&	1:36:39.560	&	+15:45:42.05	&	215780	$\pm$	12990.7	&	20.730	&	0.40	$\pm$	0.18	&	1.58	$\pm$	0.76	\\
46	&	1:36:44.298	&	+15:47:09.69	&	111911	$\pm$	15707.6	&	27.728	&	0.55	$\pm$	0.12	&	2.36	$\pm$	0.49	\\
47	&	1:36:36.203	&	+15:47:38.18	&	107064	$\pm$	24526.9	&	39.450	&	0.54	$\pm$	0.22	&	1.97	$\pm$	0.77	\\
48	&	1:36:36.647	&	+15:47:22.51	&	122600	$\pm$	7792.62	&	27.098	&	0.49	$\pm$	0.12	&	1.53	$\pm$	0.47	\\
49	&	1:36:37.832	&	+15:45:05.00	&	742744	$\pm$	14473.6	&	21.238	&	2.03	$\pm$	0.21	&	7.39	$\pm$	0.84	\\
50	&	1:36:44.693	&	+15:44:57.88	&	900466	$\pm$	37777.8	&	23.927	&	4.36	$\pm$	0.12	&	15.85	$\pm$	0.51	\\
51	&	1:36:39.115	&	+15:44:24.39	&	1310300	$\pm$	34141.2	&	38.770	&	5.77	$\pm$	0.37	&	19.48	$\pm$	1.53	\\
52	&	1:36:43.656	&	+15:44:25.81	&	134247	$\pm$	7230.15	&	12.224	&	0.49	$\pm$	0.10	&	1.43	$\pm$	0.31	\\
53	&	1:36:42.126	&	+15:44:30.09	&	194390	$\pm$	13019.4	&	21.615	&	0.04	$\pm$	0.13	&	0.05	$\pm$	0.43	\\
54	&	1:36:32.206	&	+15:45:08.55	&	1965580	$\pm$	4655.63	&		&	0.09	$\pm$	0.04	&	0.67	$\pm$	0.13	\\
55	&	1:36:37.536	&	+15:44:28.66	&	163791	$\pm$	31493.7	&	18.336	&	0.69	$\pm$	0.16	&	2.23	$\pm$	0.52	\\
56	&	1:36:34.377	&	+15:45:25.66	&	13685.8	$\pm$	9683.52	&		&	0.11	$\pm$	0.05	&	0.45	$\pm$	0.19	\\
57	&	1:36:29.639	&	+15:45:42.03	&	46417.3	$\pm$	3244.83	&	11.655	&	0.53	$\pm$	0.04	&	1.78	$\pm$	0.15	\\
58	&	1:36:47.357	&	+15:44:35.07	&	239699	$\pm$	15857.8	&		&	1.28	$\pm$	0.09	&	4.22	$\pm$	0.26	\\
59	&	1:36:46.469	&	+15:45:28.51	&	109105	$\pm$	3714.01	&	23.649	&	0.15	$\pm$	0.16	&	0.69	$\pm$	0.54	\\
60	&	1:36:45.680	&	+15:46:04.85	&	57077.9	$\pm$	3876.39	&	26.627	&	0.00	$\pm$	0.15	&	0.00	$\pm$	0.55	\\
61	&	1:36:45.482	&	+15:44:11.57	&	185256	$\pm$	7353.99	&	123.395	&	0.62	$\pm$	0.09	&	2.30	$\pm$	0.33	\\
62	&	1:36:36.648	&	+15:44:06.58	&	139346	$\pm$	18122.7	&		&	1.58	$\pm$	0.11	&	5.52	$\pm$	0.41	\\
63	&	1:36:45.582	&	+15:45:09.98	&	280236	$\pm$	20166	&	19.796	&	0.94	$\pm$	0.15	&	3.61	$\pm$	0.49	\\
64	&	1:36:41.287	&	+15:47:06.13	&	243702	$\pm$	22088.5	&	42.836	&	0.79	$\pm$	0.13	&	3.93	$\pm$	0.82	\\
65	&	1:36:42.521	&	+15:46:53.30	&	151305	$\pm$	36336.8	&	39.533	&	0.68	$\pm$	0.10	&	5.92	$\pm$	1.30	\\

\hline
\multicolumn{7}{|c|}{NGC 2976} \\
\hline
1	&	9:47:07.445	&	+67:55:51.86	&	1695260	$\pm$	60212.2	&	63.23	&	3.88	$\pm$	0.22	&	17.05	$\pm$	0.83	\\
2	&	9:47:24.622	&	+67:53:56.41	&	1082160	$\pm$	27734.4	&	38.37	&	2.45	$\pm$	0.40	&	12.83	$\pm$	0.78	\\
3	&	9:47:15.281	&	+67:55:01.27	&	289456	$\pm$	25784.9	&	58.47	&	1.90	$\pm$	0.54	&	8.65	$\pm$	1.70	\\
4	&	9:47:10.354	&	+67:55:21.94	&	60221.6	$\pm$	9778.66	&	30.12	&	0.24	$\pm$	0.24	&	0.00	$\pm$	0.80	\\
5	&	9:47:19.193	&	+67:53:57.85	&	134552	$\pm$	23128.9	&	32.58	&	1.65	$\pm$	0.68	&	5.53	$\pm$	1.90	\\
6	&	9:47:17.484	&	+67:55:47.01	&	163572	$\pm$	5770.09	&	36.58	&	0.43	$\pm$	0.18	&	3.46	$\pm$	0.59	\\
7	&	9:47:21.598	&	+67:54:46.30	&	137250	$\pm$	48280.9	&	55.72	&	1.67	$\pm$	0.43	&	6.00	$\pm$	1.58	\\

\hline
\multicolumn{7}{|c|}{NGC 3351} \\
\hline

1	&	10:43:53.263	&	+11:41:40.69	&	29756.5	$\pm$	7065.13	&	50.37	&	0.40	$\pm$	0.06	&	1.36	$\pm$	0.25	\\
2	&	10:43:53.926	&	+11:42:10.62	&	71167.1	$\pm$	9827.06	&	73.94	&	1.07	$\pm$	0.09	&	4.88	$\pm$	0.39	\\
3	&	10:43:54.888	&	+11:42:53.37	&	52865.8	$\pm$	10612.6	&	57.88	&	1.33	$\pm$	0.13	&	5.80	$\pm$	0.50	\\
4	&	10:43:59.033	&	+11:43:20.52	&	110762	$\pm$	8380.39	&	26.28	&	2.69	$\pm$	0.08	&	12.12	$\pm$	0.40	\\
5	&	10:43:57.262	&	+11:43:20.70	&	39988.9	$\pm$	7936.52	&	37.51	&	0.34	$\pm$	0.13	&	2.09	$\pm$	0.72	\\
6	&	10:44:00.686	&	+11:43:30.20	&				&	31.81	&	0.00	$\pm$	0.06	&	0.00	$\pm$	0.36	\\
7	&	10:44:03.581	&	+11:42:58.70	&				&	16.33	&	0.00	$\pm$	0.05	&	0.00	$\pm$	0.17	\\
8	&	10:44:02.822	&	+11:43:21.35	&	34441.7	$\pm$	3447.8	&	222.69	&	0.67	$\pm$	0.03	&	2.86	$\pm$	0.14	\\
9	&	10:43:55.337	&	+11:44:03.80	&	29779.4	$\pm$	6751.27	&	30.40	&	0.39	$\pm$	0.03	&	1.69	$\pm$	0.18	\\
10	&	10:44:05.335	&	+11:43:41.54	&				&	93.43	&	0.01	$\pm$	0.04	&	0.07	$\pm$	0.16	\\
11	&	10:44:04.469	&	+11:42:56.80	&	26650.2	$\pm$	1484.84	&	29.71	&	0.20	$\pm$	0.04	&	0.72	$\pm$	0.10	\\
12	&	10:44:00.619	&	+11:42:33.70	&				&	65.37	&	0.00	$\pm$	0.19	&	0.31	$\pm$	0.97	\\
13	&	10:44:01.229	&	+11:42:03.91	&	43550.1	$\pm$	3505.9	&	78.23	&	0.06	$\pm$	0.14	&	0.32	$\pm$	0.62	\\
14	&	10:43:58.224	&	+11:41:13.20	&	58854.7	$\pm$	5383.87	&	28.47	&	0.40	$\pm$	0.13	&	2.05	$\pm$	0.54	\\
15	&	10:43:56.419	&	+11:41:20.24	&	17856.3	$\pm$	1939.11	&	22.17	&	0.28	$\pm$	0.08	&	1.04	$\pm$	0.32	\\
16	&	10:43:59.585	&	+11:41:03.98	&	7658.82	$\pm$	6739.34	&	33.48	&	0.02	$\pm$	0.07	&	0.10	$\pm$	0.26	\\
17	&	10:43:53.818	&	+11:40:51.54	&	7764.99	$\pm$	4562.53	&	23.67	&	0.26	$\pm$	0.07	&	1.02	$\pm$	0.21	\\
18	&	10:43:59.131	&	+11:40:34.30	&	21423.5	$\pm$	2215.66	&	21.53	&	0.13	$\pm$	0.07	&	0.63	$\pm$	0.24	\\
19	&	10:44:03.432	&	+11:42:10.24	&	9609.7	$\pm$	3712.73	&	172.73	&	0.00	$\pm$	0.15	&	0.00	$\pm$	0.58	\\
20	&	10:44:02.414	&	+11:40:43.00	&	1287.77	$\pm$	1699.64	&	14.03	&	0.00	$\pm$	0.04	&	0.21	$\pm$	0.11	\\
21	&	10:43:59.052	&	+11:43:02.63	&	523.258	$\pm$	1028.05	&	72.59	&	0.00	$\pm$	0.16	&	0.00	$\pm$	0.60	\\
22	&	10:43:52.049	&	+11:41:49.62	&	93469.4	$\pm$	7547.26	&	14.02	&	0.29	$\pm$	0.10	&	1.40	$\pm$	0.36	\\
23	&	10:44:00.550	&	+11:41:42.28	&	26633.6	$\pm$	9863.53	&	107.27	&	0.00	$\pm$	0.11	&	0.00	$\pm$	0.51	\\
\hline



\end{longtable} 
\end{center}

\end{document}